\documentstyle[aas2pp4]{article}

\received{20 May 1998}
\accepted{29 March 1999}
\lefthead{Wanajo, Hashimoto, \& Nomoto}
\righthead{Nucleosynthesis in ONeMg Novae}

\begin{document}

\title{Nucleosynthesis in ONeMg Novae: Models versus Observations 
to Constrain the Masses of ONeMg White Dwarfs and Their Envelopes}

\author{Shinya Wanajo}
\affil{Division of Theoretical Astrophysics, 
       National Astronomical Observatory, \\
       2-21-1 Osawa, Mitaka, Tokyo 181-8588, Japan;
       shinya.wanajo@nao.ac.jp}

\author{Masa-aki Hashimoto}
\affil{Department of Physics, Faculty of Science, 
       Kyushu University, \\
       4-2-1 Ropponmatsu, Tyuo-ku, Fukuoka 810-8560, Japan;
       hashi@gemini.rc.kyushu-u.ac.jp}

\and

\author{Ken'ichi Nomoto}
\affil{Department of Astronomy \& Research Center for Early Universe, 
       School of Science, 
       University of Tokyo, \\
       7-3-1 Hongo, Bunkyo-ku, Tokyo 113-0033, Japan;
       nomoto@astron.s.u-tokyo.ac.jp}

\begin{abstract}
Nucleosynthesis in ONeMg novae has been investigated with the wide
ranges of three parameters, i.e., the white dwarf mass, the envelope
mass at ignition, and the initial composition. A quasi-analytic one-zone
approach is used with an up-to-date nuclear reaction network. The
nucleosynthesis results show correlation with the peak temperatures or
the cooling timescales during outbursts. Among the combinations of white
dwarf and envelope masses which give the same peak temperature, the
explosion is more violent for a lower white dwarf mass owing to its
smaller gravitational potential. Comparison of the nucleosynthesis
results with observations implies that at least two-third of the white
dwarf masses for the observed ONeMg novae are $\simeq 1.1 M_\odot$,
which are significantly lower than estimated by previous hydrodynamic
studies but consistent with the observations of V1974 Cyg. Moreover, the
envelope masses derived from the comparison are $\gtrsim 10^{-4}
M_\odot$, which are in good agreement with the ejecta masses estimated
from observations but significantly higher than in previous hydrodynamic
studies. With such a low mass white dwarf and a high mass envelope, the
nova can produce interesting amounts of $\gamma$-ray emitters $^7$Be,
$^{22}$Na, and $^{26}$Al. We suggest that V1974 Cyg has produced
$^{22}$Na as high as the upper limit derived from the COMPTEL survey. In
addition, a non-negligible part of the Galactic $^{26}$Al may originate
from ONeMg novae, if not the major contributors.  Both the future
INTEGRAL survey for these $\gamma$-ray emitters and abundance estimates
derived from ultraviolet, optical, and near infrared spectroscopies will
impose a severe constraint on the current nova models.
\end{abstract}


\section{INTRODUCTION}

A classical nova has been thought to be a thermonuclear runaway of
hydrogen-rich gas accumulated onto a white dwarf in a close binary
system (\cite{Trur82}; \cite{Gehr98} and references therein). Recent
observations show that about 30\% of well-studied events are classified
as oxygen-neon-magnesium (ONeMg) novae. Observationally, ONeMg novae are
characterized by strong line emissions in neon and other
intermediate-mass elements like magnesium, aluminum, silicon, and sulfur
in their ejected shells (\cite{Livi94}). The presence of these elements
implies that the accumulated gases must have been substantially enriched
through the dredge-up from the ONeMg cores.

ONeMg novae have been suggested to be a promising production site of
$\gamma$-ray emitters $^7$Be, $^{22}$Na, and $^{26}$Al (\cite{Star78};
\cite{Weis90}; \cite{Nofa91}; \cite{Star93}; \cite{Coc95};
\cite{Poli95}; \cite{Hern96}; Wanajo et al. 1997a, b; \cite{Jose97};
\cite{Jose98}; \cite{Star98}). However, the following three
uncertainties confront us when studying nucleosynthesis in ONeMg
novae. First, the mass of the ONeMg white dwarf is not constrained from
theoretical models any more than $\sim 1.1-1.4 M_\odot$, which results
from the $8 - 10 M_\odot$ stellar evolution models (Nomoto 1984, 1987;
\cite{Iben85}). On the other hand, only a few observational estimates of
the white dwarf masses have been reported (\cite{Pare95}; \cite{Krau96};
\cite{Rett97}). Second, there is a serious disagreement on the accreted
masses onto white dwarfs between observational estimates and current
theories. The ONeMg white dwarfs in previous hydrodynamic studies
accumulate a few $10^{-5} M_\odot$ of the envelope masses at ignition
(\cite{Poli95}; \cite{Star98}; \cite{Jose98}). On the other hand, the
estimated ejecta masses of QU Vul, V838 Her, and V1974 Cyg are $\sim
10^{-4} - 10^{-3} M_\odot$ (\cite{Tayl87}; \cite{Gree88}; \cite{Saiz92};
\cite{Wood92}; \cite{Pave93}; \cite{Shor93}; \cite{Saiz96};
\cite{Vanl96}; \cite{Wood97}), which are $10 - 100$ times larger than
theoretical estimates. Starrfield et al. (1998) have shown that the
envelope mass increases with decreasing mass accretion rate and white
dwarf luminosity (see also \cite{Pria95}; \cite{Kove97}). However, it is
still significantly lower than observational estimates. Third, there has
been no consensus on the mixing mechanism between the white dwarf matter
and the accreted gas, though a few hypotheses such as diffusion, shear
mixing, and convective overshooting have been proposed (\cite{Pria84};
\cite{Kutt87}; \cite{Iben91}; \cite{Glas97}; \cite{Kerc98},
b). Furthermore, the metallicity estimates for the observed ejecta of
ONeMg novae show a wide spread between 0.09 and 0.86 in mass fraction
(\cite{Livi94}; \cite{Poli95}; \cite{Star98}). The initial composition
of an envelope may significantly affect the nucleosynthesis result as
well as the energetics of the outburst (\cite{Kove97}; \cite{Jose98}).

The purpose of this study is to examine nucleosynthesis in ONeMg novae
with the wide ranges of three parameters: the white dwarf mass, the
envelope mass, and the mixing ratio of the core-surface matter into the
envelope. In \S~\ref{sec:method}, we describe our quasi-analytic nova
models and an updated nuclear reaction network. We then, in
\S~\ref{sec:comp}, compare the nucleosynthesis results for one sequence
with a previous hydrodynamic calculation.  In \S~\ref{sec:result}, we
constrain the ranges of white dwarf and envelope masses, comparing the
nucleosynthesis results with observational abundance estimates; where
the effect of changing the initial composition is considered. Finally,
the $\gamma$-ray line emissions from $^7$Be, $^{22}$Na, and $^{26}$Al
are discussed in \S~\ref{sec:gamma}.

\section{METHOD OF CALCULATION}
\label{sec:method}

\subsection{Nova Model}
\label{sec:model}

Our nova models are based on the quasi-analytic approach for the
hydrogen shell flash on a white dwarf (\cite{Sugi78}; Fujimoto 1982a,
b). The temperature and density structures of an envelope are obtained
analytically for a given set of a white dwarf mass ($M_{\rm WD}$) and an
envelope mass ($M_{\rm env}$), on the assumption that the spherical
envelope expands in hydrostatic equilibrium. We have constructed models
for 49 sets of $M_{\rm WD}$ ($1.05 - 1.35 M_\odot$) and $M_{\rm env}$
($10^{-6} - 10^{-3} M_\odot$). The former corresponds to the masses of
ONeMg cores which results from $8 - 10 M_\odot$ stellar evolutions
(Nomoto 1984, 1987; \cite{Iben85}), and the latter covers those both
from theories ($\sim 10^{-5}-10^{-4} M_\odot$; \cite{Trur77};
\cite{Poli95}; \cite{Star98}; \cite{Jose98}) and from observations
($\gtrsim 10^{-4} M_\odot$; \cite{Tayl87}; \cite{Gree88}; \cite{Saiz92};
\cite{Wood92}; \cite{Pave93}; \cite{Shor93}; \cite{Saiz96};
\cite{Vanl96}; \cite{Wood97}). The dots in Figure~\ref{fig1} are the
sequences at which our numerical calculations are performed, while
squares, triangles, and stars are taken from hydrodynamic studies by
Politano et al. (1995, hereafter PSTWS95), Starrfield et al. (1998,
hereafter STWS98), and Jos\'e \& Hernanz (1998, hereafter JH98). The
solid lines show the mass accretion rates onto the white dwarfs required
for each set of ($M_{\rm WD}$, $M_{\rm env}$), calculated by Fujimoto
(1982b). These are in reasonable agreement with those by PSTWS95,
STWS98, and JH98 ($\sim 10^{-10}-10^{-9} M_\odot$~yr$^{-1}$), but
somewhat overestimated since the luminosities of white dwarfs are
neglected and the radii are assumed to be of Chandrasekhar (see
Figure~\ref{fig3}) in Fujimoto (1982b). Note that no outburst is
achievable by an accreting white dwarf below the dashed line due to the
high accretion rate. It is obvious that a rather low accretion rate (or
a low luminosity of the white dwarf) is required to obtain a massive
envelope such as $\sim 10^{-4}-10^{-3} M_\odot$ as expected by
observations.

The quasi-analytic nova model has been elaborated by Sugimoto \&
Fujimoto (1978) and Fujimoto (1982a, b). Let us discuss the model in
some detail, since it can characterize the nova burst very well. The
pressure and the density at the base of the envelope are expressed in
terms of $M_{\rm WD}$ and $M_{\rm env}$:
\begin{eqnarray} \label{eqn:pb}
P_{\rm b}    & = & \frac{GM_{\rm WD}M_{\rm env}}{4\pi {R_{\rm WD}}^4}
                   f_{\rm b} \:, \\ \label{eqn:rhob}
\rho_{\rm b} & = & \frac{M_{\rm env}}
                   {4\pi {R_{\rm WD}}^3}V_{\rm b}f_{\rm b} \:, 
\end{eqnarray}
where $R_{\rm WD}$ is the radius of the white dwarf, $V$ is a
homologous invariant defined by
$$
V\equiv -\frac{d\ln P}{d\ln r} = \frac{GM\rho}{rP} \:.
$$
Hereafter the subscript `b' denotes a quantity at the base of the
envelope. The flatness parameter $f$ in the equations~(\ref{eqn:pb}) and
(\ref{eqn:rhob}) decreases monotonically as the shell flash proceeds:
\begin{equation}\label{eqn:f}
f\left(x,N\right) \equiv 
 \frac{x^{N+1}\left(1-x\right)^{3-N}}
      {\left(N+1\right)B_x\left(N+1,3-N\right)} \:, 
\end{equation}
where 
\begin{equation}\label{eqn:x}
x\equiv \frac{N+1}V \quad (0<x<1) \:.
\end{equation}
The value of $f_{\rm b}$ denotes the degree of the `flatness' of the
envelope. For $f_{\rm b} \sim 1$ ($x_{\rm b} \sim 0$), the envelope is
thin and strongly degenerated, and thus is flat. On the other hand, for
$f_{\rm b} \sim 0$ ($x_{\rm b} \sim 1$), the envelope is thick and non
degenerate, and thus is spherical. The polytropic index $N$ in the
equations~(\ref{eqn:f}) and (\ref{eqn:x}) is defined by $$ \frac
N{N+1}\equiv \frac{d\ln \rho}{d\ln P} \:, $$ and $B_x\left(p,q\right)$
is the incomplete beta function defined by $$ B_x\left(p,q\right)\equiv
\int_0^xt^{p-1}\left(1-t\right)^{q-1}dt \quad (0<x<1) \:.  $$
%We used the following transformation when $B_x(a,b)$ was nearly singular due 
%to $1-x \ll 1$ or $b < 0$\/:
%\begin{equation}
%B_x\left(a,b\right) 
%= B_{x/2}\left(a,b\right)
%+ B_{1-x/2}\left(b,a\right)
%- B_{1-x}\left(b,a\right)
%\end{equation}
$N$ is assumed to be adiabatic and constant throughout the envelope, but
vary with time. The effect of the spatial variation in $N$ is quite
small for a typical convective envelope (\cite{Fuji82a}). The value of
$N$ is approximately 1.5 at the beginning of a shell flash, and
approaches $\sim 3$ at the end due to the increasing radiation pressure.

The shell flash starts with $f_{\rm b} \sim 1$ ($x_{\rm b} \sim 0$).
The envelope is then heated up by nuclear burning to a thermal runaway,
and cools down when $f_{\rm b}$ decreases to $\sim 0$ ($x_{\rm b} \sim
1$). The equations (\ref{eqn:pb}) and (\ref{eqn:rhob}) are valid if
\begin{equation}
\theta\equiv \frac{Ux}{1-x}\ll 1
\label{eqn:theta}
\end{equation}
is satisfied, where $$ U\equiv \frac{d\ln M}{d\ln r} = \frac{4\pi
r^3\rho}{M} $$ is another homologous invariant. This condition is
violated only near the last phase of the shell flash ($f_{\rm b} \sim
0$). At this phase, major nuclear reactions are frozen out except for
the pp-chain, the CNO cycle, and $\beta^+$-decay. Thus, our
nucleosynthesis results may not be significantly affected.

Figure~\ref{fig2} illustrates contours for $P_{\rm b}/f_{\rm b}$ and
$\rho_{\rm b}/V_{\rm b}f_{\rm b}$ in the $M_{\rm WD}$--$M_{\rm env}$
space. These are the the proper quantities for each set of ($M_{\rm
WD}$, $M_{\rm env}$). The stronger dependence of the former on $M_{\rm
WD}$ is due to the higher power of $R_{\rm WD}$ as seen in the
equation~(\ref{eqn:pb}) and (\ref{eqn:rhob}). The temperature at the
base of the envelope $T_{\rm b}$ can be calculated by solving the
equation of state with the use of equations (\ref{eqn:pb}) and
(\ref{eqn:rhob}). The spatial variations of the pressure, the density,
and the temperature are given when the condition (\ref{eqn:theta}) is
satisfied, by
\begin{eqnarray}
P\left(x\right)
     & = & P_{\rm b}\left(\frac{x}{x_{\rm b}}\right)^{N+1}
           \left(\frac{1-x}{1-x_{\rm b}}\right)^{-(N+1)} \nonumber \\
\rho\left(x\right)
     & = & \rho_{\rm b}\left(\frac{x}{x_{\rm b}}\right)^N
           \left(\frac{1-x}{1-x_{\rm b}}\right)^{-N} \nonumber \\
T\left(x\right)
     & = & T_{\rm b}\left(\frac{x}{x_{\rm b}}\right)^{(N+1)\nabla}
           \left(\frac{1-x}{1-x_{\rm b}}\right)^{-(N+1)\nabla} \:, \nonumber
\end{eqnarray}
where $\nabla \equiv d\ln T/d\ln P$ is assumed to be adiabatic and
constant throughout the envelope, but vary with time (on the deviation
from constant $\nabla$, see Fujimoto 1982a). The value of $x$ decreases
monotonically with increasing radius, approaching zero at the surface of
the envelope. The surface radius $R$ is given when the condition
(\ref{eqn:theta}) is satisfied, by
\begin{equation} \label{eqn:r}
R = \frac{R_{\rm WD}}{1 - x_{\rm b}}.
\end{equation}
%The mass coordinate can be written as a function of $x$:
%\begin{equation}
%M\left(x\right) = M_{\rm env}\left[1-\frac{B_x\left(N+1,3-N\right)}
%                        {B_{x_{\rm b}}\left(N+1,3-N\right)}\right] \:,
%\end{equation}
%where the origin is taken at the base $x = x_{\rm b}$.
Now we know the envelope structure completely.

The progress of a shell flash is derived by energy conservation,
\begin{equation} \label{eqn:dsdt}
\frac{ds}{dt}=\frac{\varepsilon_{\rm N}}
                   {\left\langle T\right\rangle} \:,
\end{equation}
where $\varepsilon_{\rm N}$ is the nuclear energy generation rate per
unit mass, $s$ is the specific entropy which is spatially constant in
the convective envelope, and $\left\langle T \right\rangle$ is the mass
averaged temperature over the envelope.
%\begin{eqnarray}
%\left\langle T\right\rangle & = & 
% T_{\rm b}\left(\frac{1-x_{\rm b}}{x_{\rm b}}\right)^{(N+1)\nabla}
% \nonumber \\
% & & \times
% \frac{B_{x_{\rm b}}\left(N+(N+1)\nabla+1,3-(N+1)\nabla-N\right)}
%      {B_{x_{\rm b}}\left(N+1,3-N\right)} \:. 
%\end{eqnarray}
The energy inflow from the white dwarf and loss from the photosphere are
neglected, being much smaller than the nuclear energy during the
explosive hydrogen burning. The time variation of $x_{\rm b}$ is then
calculated from the equations~(\ref{eqn:pb}), (\ref{eqn:rhob}), and
(\ref{eqn:dsdt}) with the use of the equation of state. The expansion
velocity of the envelope $v_{\rm exp}$ is derived from the
equation~(\ref{eqn:r}) as $$ v_{\rm exp} = \frac{R}{1 - x_{\rm
b}}\frac{dx_{\rm b}}{dt}.  $$ Each calculation is started with the
initial temperature $T_{\rm b} = 5 \times 10^7$~K, and ceased when the
nuclear luminosity decreases to the Eddington luminosity where no
further heavy elements are synthesized.

The $M_{\rm WD}$--$R_{\rm WD}$ relation is derived for an isothermal
core ($2 \times 10^7$~K) consists of oxygen, neon ($ = 5 : 3$), and
partially degenerate electron gases including the effect of the Coulomb
interaction (\cite{Ichi94}), as shown in Figure~\ref{fig3}. The
solid line denotes our results and the triangles are taken from PSTWS95
and STWS98. Our results are between those of carbon and magnesium white
dwarfs by Hamada \& Salpeter (1961), and somewhat smaller than by
PSTWS95 and STWS98. A variation of $R_{\rm WD}$ significantly influences
the density due to $\rho_{\rm b} \propto {R_{\rm WD}}^{-3}$ as seen in
the equation~(\ref{eqn:rhob}), much more than the temperature ($\propto
{R_{\rm WD}}^{-1}$). Note that the ONe white dwarf is unable to increase
its mass beyond $1.38 M_\odot$ because the electron capture on $^{20}$Ne
and $^{24}$Mg triggers the collapse (denoted by a dot on the solid line;
Nomoto 1984, 1987).

\subsection{Nuclear reaction network and initial composition}
\label{sec:ntwk}

The nuclear reaction network used in this work contains 87 stable and
proton-rich isotopes from hydrogen to calcium (Table~\ref{tab:ntwk}),
including all relevant nuclear reactions and weak interactions. The
reaction $^8$B($p$, $\gamma$)$^9$C, which can be a sink for the $^7$Be
production (\cite{Boff93}), is also included. The ground and isomeric
states of $^{26}$Al take longer than the mean lifetime of the isomer
($\simeq 9.2$~s) to be equilibrated for $\lesssim 4 \times 10^8$~K
(\cite{Ward80}). The peak temperatures in the models responsible for the
observed ONeMg novae may be less than $4 \times 10^8$~K as will be
discussed in \S~\ref{sec:obs}. Thus, the two states are separated as
different isotopes. The nuclear reaction rates are taken from Thielemann
et al. (1995). They are based on the rates by Caughlam \& Fowler (1988),
those calculated by a statistical model (\cite{Trur87}), and the latest
experimental data (\cite{VanW94}, etc.). We also include new reaction
rates by Herndl et al. (1995) and Iliadis et al. (1996). The rate
$^{26}$Si($p$, $\gamma$)$^{27}$P (\cite{Hern95}) may have a special
importance, being $10^3 - 10^4$ times larger than the previous one in
the typical nova temperature range. The rates $^{25}$Mg($p$,
$\gamma$)$^{26}$Al and $^{25}$Al($p$, $\gamma$)$^{26}$Si (\cite{Ilia96})
may be also of importance for $^{26}$Al production, though the latter
involves a large uncertainty. In our computations, all nuclear reaction
rates are mass-averaged over the envelope except for $\beta^+$-decay
which does not depend on density and temperature.
%as 
%\begin{eqnarray}
%\frac{1}{{\rho_{\rm b}}^{k-1}}\left\langle\rho^{k-1}\lambda\right\rangle
%&=& \frac{1}{{\rho_{\rm b}}^{k-1}M_{\rm env}}
%    \int^{M_{\rm WD}+M_{\rm env}}_{M_{\rm WD}}\rho^{k-1}\lambda\,dM
%    \nonumber\\
%&=& \frac{\left[\left(1-x_{\rm b}\right)/x_{\rm b}\right]^{kN}}
%         {B_{x_{\rm b}}\left(N+1,3-N\right)}
%    \int^{x_{\rm b}}_{0}\lambda\left[T\left(x\right)\right]
%                        x^{kN}\left(1-x\right)^{2-kN}\,dx\:, 
%\end{eqnarray}
%for $k$-body reactions, 
%where $\lambda\equiv N_{\rm A}\left\langle\sigma v\right\rangle$. 

\begin{table}[b]
\caption{Nuclear Reaction Network Employed}
\label{tab:ntwk}
\smallskip
\begin{tabular}{rrrrrr}
\hline
\hline
Element & $A_{\rm min}$ & $A_{\rm max}$ &
Element & $A_{\rm min}$ & $A_{\rm max}$ \\
\hline
H \dotfill &  1  &  2  & Na\dotfill & 20 & 23 \\
He\dotfill &  3  &  4  & Mg\dotfill & 21 & 26 \\
Li\dotfill &  7  &  7  & Al\dotfill & 22 & 27 \\
Be\dotfill &  7  &  7  & Si\dotfill & 24 & 30 \\
B \dotfill &  8  & 11  & P \dotfill & 27 & 31 \\
C \dotfill &  9  & 13  & S \dotfill & 28 & 34 \\
N \dotfill & 13  & 15  & Cl\dotfill & 31 & 37 \\
O \dotfill & 14  & 18  & Ar\dotfill & 32 & 38 \\
F \dotfill & 17  & 19  & K \dotfill & 35 & 39 \\
Ne\dotfill & 18  & 22  & Ca\dotfill & 36 & 40 \\
\hline
\end{tabular}
\end{table}

\begin{table}[b]
%\begin{center}
\caption{Abundances of the ONeMg Core at the Surface}
\label{tab:onemg}
\smallskip
\begin{tabular}{rrrr}
\hline
\hline
Nucleus & Mass Fraction & Nucleus & Mass Fraction \\
\hline
$^{12}$C   & 3.95E-02      & $^{24}$Mg  & 4.20E-02      \\
$^{16}$O   & 5.42E-01      & $^{25}$Mg  & 6.29E-03      \\
$^{20}$Ne  & 3.31E-01      & $^{26}$Mg  & 4.57E-03      \\
$^{21}$Ne  & 2.87E-03      & $^{27}$Al  & 1.25E-02      \\
$^{22}$Ne  & 1.34E-03      & $^{28}$Si  & 2.46E-03      \\
$^{23}$Na  & 1.65E-02      &                            \\
\hline
\end{tabular}
%\end{center}
\end{table}

The initial composition of an envelope is assumed to be a mixture of the
solar composition gas and the dredged-up matter from the surface of the
ONeMg white dwarf. The solar abundances are adopted from \cite{Ande89},
and the abundances of the ONeMg core matter from \cite{Hash93} for the
1.35 $M_\odot$ ONeMg core (Table~\ref{tab:onemg}). As can be seen in
Table~\ref{tab:onemg}, ${\rm O} : {\rm Ne} : {\rm Mg} \approx 10 : 6 :
1$, which is in good agreement with those in Nomoto and Hashimoto (1988)
for $M_{\rm WD} = 1.26, 1.36 M_\odot$ and Ritossa, Garc\'{\i}a, \& Iben
(1996) for $M_{\rm WD} = 1.2 M_\odot$.  This implies that the
composition of an ONeMg core does not significantly depend on its
mass. The mass fraction of the dredge-up matter from the ONeMg core in
the envelope $X_{\rm WD}$, which is the third parameter in this study,
is of importance on the nucleosynthesis results as will be discussed in
\S~\ref{sec:depz}. However, abundance estimates in the observations of
nova ejecta involve large uncertainties as pointed out by Livio \&
Truran (1994). The estimated metallicities of the six observed ONeMg
nova ejecta range widely (see Table~\ref{tab:obs}) and, unfortunately,
different authors have provided different values even for the identical
events (\cite{Will85}; \cite{Snij87}; \cite{Saiz92}; \cite{Andr94};
\cite{Aust96}; \cite{Saiz96}; \cite{Vanl96}; \cite{Vanl97}). In
addition, no consensus has been achieved in theoretical modeling how and
when the core-matter mixes into the envelope (\cite{Pria84};
\cite{Iben91}; \cite{Kutt87}; \cite{Glas97}; \cite{Kerc98}, b). Thus, we
examine all the combinations of ($M_{\rm WD}$, $M_{\rm env}$) for $X_{\rm
WD} = 0.1$ (case A), 0.4 (case B), and 0.8 (case C), which cover
observational uncertainties in abundance determinations. The initial
compositions for each case are given in Table~\ref{tab:init}.

\begin{table*}[p]
\begin{center}
%\caption{Initial Compositions of the Envelope by Mass}
Table 3: Initial Compositions of the Envelope by Mass \\
\label{tab:init}
\medskip
\begin{tabular}{rrrr}
\hline
\hline
$X_{\rm WD}$ & 0.1 & 0.4 & 0.8 \\
\hline
$p$       & 6.36E-01  & 4.24E-01 & 1.41E-01 \\
D         & 4.33E-05  & 2.88E-05 & 9.62E-06 \\
$^{ 3}$He & 2.64E-05  & 1.76E-05 & 5.87E-06 \\
$^{ 4}$He & 2.48E-01  & 1.65E-01 & 5.51E-02 \\
$^{ 7}$Li & 8.43E-09  & 5.62E-09 & 1.87E-09 \\
$^{11}$B  & 4.26E-09  & 2.84E-09 & 9.46E-10 \\
$^{12}$C  & 6.69E-03  & 1.76E-02 & 3.22E-02 \\
$^{13}$C  & 3.29E-05  & 2.19E-05 & 7.31E-06 \\
$^{14}$N  & 9.96E-04  & 6.64E-04 & 2.21E-04 \\
$^{15}$N  & 3.93E-06  & 2.62E-06 & 8.74E-07 \\
$^{16}$O  & 6.28E-02  & 2.22E-01 & 4.35E-01 \\
$^{17}$O  & 3.50E-06  & 2.34E-06 & 7.79E-07 \\
$^{18}$O  & 1.95E-05  & 1.30E-05 & 4.34E-06 \\
$^{19}$F  & 3.65E-07  & 2.43E-07 & 8.11E-08 \\
$^{20}$Ne & 3.45E-02  & 1.33E-01 & 2.65E-01 \\
$^{21}$Ne & 2.90E-04  & 1.15E-03 & 2.29E-03 \\
$^{22}$Ne & 2.51E-04  & 6.15E-04 & 1.10E-03 \\
$^{23}$Na & 1.68E-03  & 6.60E-03 & 1.32E-02 \\
$^{24}$Mg & 4.66E-03  & 1.71E-02 & 3.37E-02 \\
$^{25}$Mg & 6.89E-04  & 2.55E-03 & 5.04E-03 \\
$^{26}$Mg & 5.27E-04  & 1.88E-03 & 3.67E-03 \\
$^{27}$Al & 1.31E-03  & 5.05E-03 & 1.00E-02 \\
$^{28}$Si & 8.34E-04  & 1.38E-03 & 2.10E-03 \\
$^{29}$Si & 3.09E-05  & 2.06E-05 & 6.86E-06 \\
$^{30}$Si & 2.12E-05  & 1.41E-05 & 4.71E-06 \\
$^{31}$P  & 7.35E-06  & 4.90E-06 & 1.63E-06 \\
$^{32}$S  & 3.57E-04  & 2.38E-04 & 7.93E-05 \\
$^{33}$S  & 2.90E-06  & 1.94E-06 & 6.45E-07 \\
$^{34}$S  & 1.68E-05  & 1.12E-05 & 3.74E-06 \\
$^{35}$Cl & 2.28E-06  & 1.52E-06 & 5.07E-07 \\
$^{37}$Cl & 7.70E-07  & 5.13E-07 & 1.71E-07 \\
$^{36}$Ar & 6.98E-05  & 4.65E-05 & 1.55E-05 \\
$^{38}$Ar & 1.39E-05  & 9.24E-06 & 3.08E-06 \\
$^{39}$K  & 3.13E-06  & 2.08E-06 & 6.95E-07 \\
$^{40}$Ca & 5.40E-05  & 3.60E-05 & 1.20E-05 \\
\hline
\end{tabular}
\end{center}
\end{table*}
\addtocounter{table}{1}

\section{Comparison with nucleosynthesis by a hydrodynamic model}
  \label{sec:comp}

Up to now, a number of works on nucleosynthesis in ONeMg novae have been
performed (\cite{Hill82}; \cite{Weis90}; \cite{Nofa91}, and references
therein). Their nova models were, however, based on one-zone envelopes,
using the spatially constant temperature and density profiles taken from
hydrodynamic studies (\cite{Star78}; \cite{Star88}). Coc et al. (1995)
have studied $^{22}$Na and $^{26}$Al production in ONeMg novae with
another semi-analytic method (\cite{MacD83}). Their nova model and ours
give similar envelope structures in temperature and density. However,
our model includes the effect of the partially degenerate and
relativistic electron gas, while Coc et al. (1995) treated electrons as
the ideal gas. The electron degeneracy can not be neglected in the early
phase of outbursts. Hernanz et al. (1996) and Jos\'e, Hernanz, \& Coc
(1997) have also examined nucleosynthesis in novae with the use of a
hydrodynamic method. However, they focused on $^7$Li or $^{26}$Al
production, and gave only a few synthesized isotopes in their papers.

Hence, we compare our model with sequence~6 in STWS98 to see the
differences of nucleosynthesis between the quasi-analytic and
hydrodynamic methods. The nova model in STWS98 was identical to that of
PSTWS95, except that the former included the updated nuclear reaction
rates (\cite{VanW94}; \cite{Hern95}) and OPAL opacity tables
(\cite{Igle93}). In addition, STWS98 employed a lower white dwarf
luminosity and a lower mass accretion rate to obtain a more massive
ignition envelope. Furthermore, an important change was that STWS98 used
a longer mixing length of $(2 - 3) \times$~pressure scale height. We do
not compare our results with JH98 which has studied nucleosynthesis in
ONeMg (and CO) novae using a hydrodynamic code, since the white dwarf
radii are not presented. Their results showed, however, similar trends
to PSTWS95 and STWS98. We use the same initial composition, the nuclear
reaction rates, $M_{\rm WD}$ ($ = 1.25 M_\odot$), $M_{\rm env}$ ($ = 4.5
\times 10^{-5} M_\odot$), and $R_{\rm WD}$ as STWS98 for
comparison. Note that the nucleosynthesis results in this work are
obtained for the whole envelope, while those in STWS98 for the ejected
matter. Thus, STWS98 may strongly reflect the composition of the outer
region. Figure~\ref{fig4} shows the ratios of isotopes (dots) and
elements (triangles) between ours and STWS98 (sequence 6). Our
calculation obtains a higher peak temperature ($\simeq 3.17 \times
10^8$~K) than that of STWS98 ($\simeq 3.00 \times 10^8$~K), since the
latter model ignited hydrogen one zone above the base so that the
envelope is effectively thinner (see STWS98). The prominent
underproduction of several isotopes like $^{15}$N, $^{18}$O, $^{21}$Ne,
$^{22}$Na (and perhaps $^{23}$Na not shown in STWS98; see PSTWS95 for
instance), $^{24}$Mg, and $^{26}$Mg is due to our assumption of a fully
convective one-zone envelope. Since these isotopes are rather fragile
against the ($p$, $\gamma$) or ($p$, $\alpha$) reactions, they decrease
significantly even at the late phase of the outburst. In contrast, these
isotopes were able to survive in STWS98, escaping from the hotter
convective region into the cooler radiative region at the late
phase. Especially, $^{15}$N is extremely fragile against the ($p$,
$\alpha$) reaction, thus being underproduced by more than 5 orders of
magnitude in this work. As a result, nitrogen (mostly $^{14}$N in this
work) is also underproduced compared to STWS98 in which $^{15}$N is
dominant. On the other hand, carbon ($^{12}$C and $^{13}$C) are
significantly overproduced, transferred from $^{15}$N. We should be
careful on these differences in comparing the nucleosynthesis results
with observations. However, both results are in excellent agreement for
other isotopes, and especially for elements (except for carbon and
nitrogen) which are more important for comparison with observations.

\section{Nucleosynthesis in ONeMg novae}
\label{sec:result}

\subsection{Nuclear flows in the $N$--$Z$ plane}
\label{sec:nzpl}

In this section, we present some important aspects of nucleosynthesis in
ONeMg novae, referring to the results of several ($M_{\rm WD}$, $M_{\rm
env}$) models. Figure~\ref{fig5} shows the final abundances and
the net nuclear flows in the $N$--$Z$ plane. The size of a circle
denotes the mole fraction of the isotope defined by $Y_i \equiv X_i/A_i$
in the logarithmic scale. The initial composition is shown by dotted
circles. The net nuclear flow of a reaction from the $i$-th to $j$-th
isotope, defined as
$$
F_{ij} \equiv
 \int \left[ \dot Y_i\left( i\rightarrow j\right) 
             -\dot Y_j\left(j\rightarrow i\right) \right] dt \:,
$$
is denoted by the length of an arrow in the same scale. The mixing ratio
$X_{\rm WD}$ is assumed to be 0.4 (case~B) throughout this section,
which is close to the average metallicity of the ejecta estimated from
observations (see `$Z$' in Table~\ref{tab:obs}).

Figure~\ref{fig6} shows the peak temperature at the base $T_{\rm
peak}$, the cooling timescale $\tau$ defined as the duration from the
peak to its half in temperature, the peak nuclear energy generation rate
per unit mass $\varepsilon_{\rm peak}$, and the ejection velocity
$v_{\rm ej}$ in the $M_{\rm WD}$--$M_{\rm env}$ space. Here, $v_{\rm
ej}$ is defined as the expansion velocity $v_{\rm exp}$ when it equals
the escape velocity $v_{\rm esc}$ (for the models denoted by
circles). For the models denoted by crosses in which $v_{\rm exp}$ is
below $ v_{\rm esc}$ throughout the calculations, $v_{\rm ej}$ is
replaced with $v_{\rm exp}$ at the maximum.  As seen in
Figure~\ref{fig6}, $\tau$ has a weaker dependence on $M_{\rm WD}$
than $T_{\rm peak}$, while the trend of $\varepsilon_{\rm peak}$ is
similar to $T_{\rm peak}$. As a result, among the models of the same
peak temperature, the explosion is more violent for the smaller $M_{\rm
WD}$ due to its smaller gravitational potential. This is also seen in
the panel of $v_{\rm ej}$, which shows the similar trend to $\tau$ in
the $M_{\rm WD}$--$M_{\rm env}$ space. In order to obtain the fast
ejection velocities such as $\gtrsim 1000$~km~s$^{-1}$ as derived by
recent observations (\cite{Gehr98} and references therein), the cooling
timescale must be $\lesssim 1000$~s where the $\beta^+$-decay of
$^{14}$O ($\tau \simeq 102$~s) and $^{15}$O ($\tau \simeq 176$~s) plays
an important role.

\subsubsection{Low temperature sequences}

For the model ($M_{\rm WD}/M_\odot$, $M_{\rm env}/M_\odot$) = (1.10,
$10^{-4.5}$), the initially present $^{24}$Mg is entirely transferred to
silicon, even though $T_{\rm peak}$ is as low as $\sim 2 \times 10^8$~K
(Figure~\ref{fig5}). In contrast, the initial $^{20}$Ne remains
mostly unburnt, though minor nuclear flows appear through the Ne-Na
cycle. A part of the initial $^{16}$O is converted to $^{17}$O,
$^{12}$C, $^{13}$C, and $^{14}$N. The HCNO cycle is active near the peak
in temperature, turning to the CNO cycle as the temperature
decreases. Thus, almost all $^{15}$N is eventually converted to
$^{14}$N, $^{12}$C, and $^{13}$C. Note that, for the models with $T_{\rm
peak} \lesssim 2 \times 10^8$~K, $v_{\rm exp}$ is too small to overcome
$v_{\rm esc}$ as seen in Figure~\ref{fig6}.

\subsubsection{Moderate temperature sequences}

The nucleosynthesis results for ($M_{\rm WD}/M_\odot$, $M_{\rm
env}$ $/M_\odot$) = (1.15, $10^{-4.0}$) and (1.35, $10^{-5.5}$) (hereafter
N1540B and N3555B, respectively) differ significantly, regardless of
their mostly same $T_{\rm peak}$ ($\simeq 2.9 \times 10^8$~K) as seen in
Figure~\ref{fig5}. This can be explained as follows.
Figure~\ref{fig7} shows the time variations of $T_{\rm b}$ and
$\varepsilon$ for each model. The cooling timescale for N1540B ($\tau
\simeq 190$~s) is more than one order shorter than for N3555B ($\tau
\simeq 2400$~s). This is a consequence of the weaker gravitational
potential for N1540B owing to its smaller $M_{\rm WD}$ and thus its
larger $R_{\rm WD}$ (Figure~\ref{fig3}). In addition, the nuclear
energy generation rate remains as high as $\sim
10^{14}$~erg~g$^{-1}$~s$^{-1}$ even after the envelope expands and the
temperature decreases to $\sim 10^8$~K, owing to the $\beta^+$-decay of
$^{14}$O, $^{15}$O, and other unstable nuclei. As a result, the
expansion of the envelope is accelerated and then the temperature drops
fairly quickly, even when its structure is returning to the static
configuration. In contrast, for N3555B, almost all the short-lived
$\beta^+$-unstable nuclei have decayed at the late phase. Hence, the
temperature drops slowly with the decreasing nuclear energy generation
rate. The patterns of the temperature decreases are, therefore, not
similar between these models. The critical cooling timescale between the
slow (N3555B) and fast (N1540B) expansion is $\tau \sim 1000$~s. The
cooling timescale for N1540B is comparable to the $\beta^+$-decay
lifetime of $^{15}$O (=~176~s). As a result, $^{15}$N survives the
following ($p$, $\alpha$) reactions and significantly enhances. For
similar reasons, $^{18}$O, $^{25}$Mg, and $^{26}$Al are prominent in
N1540B, while they are absent in N3555B. Note that the somewhat higher
$\varepsilon_{\rm peak}$ in N1540B is due to the higher density at the
base (Figure~\ref{fig2}).

It is noteworthy that the net nuclear flows of $^{24}$Mg($p$,
$\gamma$)$^{25}$Al have overcome the initial abundance of $^{24}$Mg for
both N1540B and N3555B (Figure~\ref{fig5}), owing to substantial
nuclear flux from the Ne-Na region. It implies that the initial amount
of $^{24}$Mg does not significantly affect the production of isotopes $A
\ge 24$ for the models $T_{\rm peak} \gtrsim 3 \times 10^8$~K. Note that
N1540B also obtains the significantly higher ejection velocity ($\simeq
2100$~km~s$^{-1}$) than N3555B ($\simeq 1200$~km~s$^{-1}$). As seen in
Figure~\ref{fig6}, for all the models with $T_{\rm peak} \gtrsim 3
\times 10^8$~K, $v_{\rm exp}$ exceeds $v_{\rm esc}$ and obtains $v_{\rm
ej} \gtrsim 1000$~km~s$^{-1}$, which is in good agreement with recent
observations of ONeMg novae.

\subsubsection{High temperature sequences}

For the models ($M_{\rm WD}/M_\odot$, $M_{\rm env}/M_\odot$) = (1.20,
$10^{-4.0}$) and (1.20, $10^{-3.5}$) (hereafter N2040B and N2035B,
respectively), substantial nuclear fluxes appear in the Ne-Na region
because of their high $T_{\rm peak}$ ($\simeq 3.3 \times 10^8$~K and
$4.2 \times 10^8$~K, respectively) as seen in
Figure~\ref{fig5}. In addition, various nuclear paths open in the
Mg-S region. The abundance of $^{26}$Al is highly enhanced in N2040B due
to the substantial nuclear flux from the Ne-Na region via $^{23}$Na($p$,
$\gamma$)$^{24}$Mg. On the other hand, $^{26}$Al is less abundant in
N2035B because of its higher peak temperature. Instead, $^{18}$O,
$^{22}$Na, and $^{23}$Na are highly enhanced in N2035B, since $\tau$
($\simeq9.5$~s) is comparable to the $\beta^+$-decay lifetimes of
$^{18}$Ne (2.4~s), $^{22}$Mg (5.6~s), and $^{23}$Mg (16~s). For the
extremely high temperature ($T_{\rm peak} \simeq 7.3 \times 10^8$~K)
model ($M_{\rm WD}/M_\odot$, $M_{\rm env}/M_\odot$) = (1.30,
$10^{-3.0}$), almost all the initial $^{20}$Ne is burned out, and the
nuclear flow extends to calcium by the rp-process
(Figure~\ref{fig5}). The leakage from the CNO cycle via the
$\alpha$-capture of $^{14}$O and $^{15}$O appears, though its
contribution to the heavy element production is negligible.

\subsection{Element and isotope production}
\label{sec:mmpl}

{\scriptsize
\begin{table*}[t]
\caption{Observed ONeMg Nova Abundances}
\label{tab:obs}
\smallskip
\begin{tabular}{rlllllllllll}
\hline
\hline
  & H & He & C & N & O & Ne & Mg & Al & Si & S & $Z$ \\
\hline
V693 CrA\tablenotemark{1}  & 2.8E-01 & 3.2E-01 & 5.1E-03 & 8.4E-02 & 1.2E-01 & 1.7E-01 & 7.6E-03 & 3.4E-03 & 2.6E-03 &         & 4.0E-01 \\
V693 CrA\tablenotemark{2}  & 1.6E-01 & 1.8E-01 & 7.9E-03 & 1.4E-01 & 2.1E-01 & 2.7E-01 & 1.8E-02 &         & 6.9E-03 &         & 6.5E-01 \\
V693 CrA\tablenotemark{3}  & 3.9E-01 & 2.0E-01 & 4.3E-03 & 8.0E-02 & 7.5E-02 & 2.3E-01 & 2.9E-03 & 1.9E-03 & 8.7E-03 &         & 4.1E-01 \\
V1370 Aql\tablenotemark{4} & 4.9E-02 & 8.8E-02 & 3.5E-02 & 1.4E-01 & 5.1E-02 & 5.2E-01 & 6.8E-03 &         & 1.8E-03 & 1.0E-01 & 8.6E-01 \\
V1370 Aql\tablenotemark{2} & 4.5E-02 & 1.0E-01 & 5.0E-02 & 1.9E-01 & 3.7E-02 & 5.6E-01 & 7.9E-03 &         & 4.6E-03 &         & 8.5E-01 \\
QU Vul\tablenotemark{5}    & 3.0E-01 & 6.0E-01 & 1.0E-03 & 2.1E-02 & 1.6E-02 & 2.3E-02 & 1.7E-03 &         & 4.0E-02 &         & 1.0E-01 \\
QU Vul\tablenotemark{2}    & 3.3E-01 & 2.7E-01 & 9.6E-03 & 7.4E-02 & 1.8E-01 & 8.7E-02 & 3.7E-03 & 9.9E-03 & 3.2E-02 & 1.2E-02 & 4.0E-01 \\
V351 Pup\tablenotemark{6}  & 3.8E-01 & 2.4E-01 & 5.9E-03 & 7.4E-02 & 1.9E-01 & 1.1E-01 &         & 4.3E-03 & 1.9E-03 &         & 3.8E-01 \\
V838 Her\tablenotemark{3}  & 6.0E-01 & 3.1E-01 & 1.2E-02 & 1.4E-02 & 2.5E-03 & 5.8E-02 &         &         &         & 2.8E-03 & 9.0E-02 \\
V1974 Cyg\tablenotemark{7} & 1.8E-01 & 3.1E-01 & 5.4E-02 & 7.7E-02 & 2.7E-01 & 1.1E-01 &         &         &         &         & 5.1E-01 \\
\hline
\end{tabular}

\smallskip
\noindent
References: $^1$Williams et al. 1985, $^2$Andre\"a, Drechsel, 
\& Starrfield 1994, $^3$Vanlandingham, Starrfield, \& Shore 1997, 
$^4$Snijder et al. 1987, $^5$Saizar et al. 1992, $^6$Saizar et al. 1996, 
$^7$Austin et al. 1969
\end{table*}
}

In this section, we discuss the global trends of element production and
isotope ratios in the $M_{\rm WD}$--$M_{\rm env}$ space, referring to
the abundances of ONeMg nova ejecta estimated from recent
observations. Table~\ref{tab:obs} shows the abundances for the recent
six ONeMg novae, V693 CrA (\cite{Will85}; \cite{Andr94}; \cite{Vanl97}),
V1370 Aql (\cite{Snij87}; \cite{Andr94}), QU Vul (\cite{Saiz92};
\cite{Andr94}), V351 Pup (\cite{Saiz96}), V838 Her (\cite{Vanl97}), and
V1974 Cyg (\cite{Aust96}). Note that the abundances of the elements not
presented in the above references are assumed to be zero, thus involving
errors by a few percent. The average metallicity for these ONeMg novae
is $\simeq 0.43$ by mass. The mixing ratio $X_{\rm WD}$ is, therefore,
assumed to be 0.4 (case~B) throughout this section. However, V1370 Aql
and V838 Her show significantly different metallicities from case~B. The
dependence on the initial composition is discussed in \S~\ref{sec:depz}.

When temperature is higher than $\sim 2 \times 10^8$~K, proton captures
are fast enough to compete with the $\beta^+$-decay of various unstable
isotopes. As a result, the nucleosynthesis results are significantly
deviated from those in steady nuclear flows like the CNO and Ne-Na
cycles. Figures~\ref{fig8}--\ref{fig14} show the final
abundances and isotope ratios by mass in the $M_{\rm WD}$--$M_{\rm env}$
space. The abundances are shaded from white (0.1) to black ($10^{-5}$)
in the logarithmic scale (except for beryllium and boron). In the rest
of this paper, all abundances are given in mass fraction. As described
below, we find that there exist two types of elements, namely, those
correlated to $T_{\rm peak}$ (e.g., oxygen, neon, and sulfur) and to
$\tau$ (e.g., carbon, sodium, and magnesium).

\subsubsection{Beryllium and Boron}
\label{sec:beb}

As seen in Figure~\ref{fig8}, the abundance of $^7$Be (in mass
fraction) reaches $\sim 10^{-6}$ for $T_{\rm peak} \sim 2.5 - 4 \times
10^8$~K (Figure~\ref{fig6}), by the $\alpha$-capture of the
initially present $^3$He. For the same $T_{\rm peak}$, the lower $M_{\rm
WD}$ models produce more $^7$Be than higher ones. This is due to the
higher densities for the formers as seen in Figure~\ref{fig2}. When
density is less than $\sim 10^3$~g~cm$^{-3}$ at temperature $\sim 2 - 4
\times 10^8$~K, the proton capture of $^7$Be is suppressed by its
inverse reaction (\cite{Boff93}). For $T_{\rm peak} \gtrsim 4 \times
10^8$~K, $^7$Be decreases by its $\alpha$-capture. As a result, the
abundance of $^{11}$B reaches $\sim 10^{-7}$. For $T_{\rm peak} \gtrsim
6 \times 10^8$~K, the abundance of $^{11}$B decreases owing to the
reaction $^{11}$C($\alpha$, $p$).

\subsubsection{Carbon and nitrogen}
\label{sec:cn}

In the steady flow of the CNO cycle ($\lesssim 2 \times 10^8$~K), the
most abundant isotope is $^{14}$N and the isotope ratios are determined
by the nuclear reaction rates as
\begin{eqnarray}
^{12}{\rm C}/^{13}{\rm C}
& = & \lambda\left[^{13}{\rm C}(p,\gamma)\right]/
\lambda\left[^{12}{\rm C}(p,\gamma)\right]\sim 2-4 \nonumber \\
^{14}{\rm N}/^{15}{\rm N}
& = & \lambda\left[^{15}{\rm N}(p,\alpha)\right]/
\lambda\left[^{14}{\rm N}(p,\gamma)\right] \nonumber \\
& \sim & 5000-50000. \nonumber
\end{eqnarray}
When temperature exceeds $\sim 2 \times 10^8$~K, the CNO cycle is
replaced with the HCNO cycle via $^{13}$N($p$,
$\gamma$)$^{14}$O($\beta^+\nu$)$^{14}$N. The abundance patterns of the
carbon and nitrogen (Figure~\ref{fig9}) mainly depend on $\tau$
(Figure~\ref{fig6}) as follows: (1) For $\tau \gg 1000$~s, the
carbon and nitrogen isotopes show the typical feature of the steady CNO
cycle, i.e., ${\rm C/N} \ll 1$, $^{12}{\rm C}/^{13}{\rm C} \sim 3$, and
$^{14}{\rm N}/^{15}{\rm N} \sim 30000$. (2) For $\tau \sim 1000$~s,
however, these isotope ratios approach $\sim 1$, due to the
$\beta^+$-decay lifetimes of $^{13}$N ($\simeq 862$~s) and $^{15}$O
($\simeq 176$~s) comparable to the cooling timescale. The thermonuclear
runaway ceases before most $^{13}$N (and some $^{15}$O) decays, and thus
the ratio C/N also reaches $\sim 1$. (3) For $\tau \ll 1000$~s, the
thermonuclear runaway ceases during the active HCNO cycle where $^{14}$O
and $^{15}$O are abundant, resulting in ${\rm C/N} \ll 1$. The ratio
$^{12}{\rm C}/^{13}{\rm C}$ is unchanged ($\sim 3$), while $^{14}{\rm
N}/^{15}{\rm N}$ is significantly reduced to $\sim 0.1$.

The abundance of nitrogen is $\sim 0.1$ in the whole area of the $M_{\rm
WD}$--$M_{\rm env}$ space, regardless of the ratio $^{14}{\rm
N}/^{15}{\rm N}$ ranging over 5 orders of magnitude. In contrast, the
abundance of carbon ranges widely ($\sim 0.001 - 0.1$) reaching its
maximum at $\tau \sim 1000$~s, while the ratio $^{12}{\rm C}/^{13}{\rm
C}$ is not significantly changed in the $M_{\rm WD}$--$M_{\rm env}$
space. The above results explain the abundance feature of the recent
ONeMg novae (Table~\ref{tab:obs}), in which the abundance of carbon
spreads widely ($\sim 0.001 - 0.01$) while that of nitrogen is $\sim
0.1$. Note that the abundance of nitrogen for QU Vul (\cite{Saiz92}) and
V838 Her (\cite{Vanl96}) is as low as $\sim 0.02$, owing to the
significantly lower metallicities ($\sim 0.1$). For V838 Her and V1974
Cyg, the ratio of C/N is $\sim 1$ which is obtained by the models with
$\tau \sim 1000$~s.

It should be noted that our models may significantly underproduce
$^{15}$N, that causes the too large ratio C/N as discussed in
\S~\ref{sec:comp}. This may be, however, only the case in the models
$\tau \gg 1000$~s. For the models $\tau \lesssim 1000$~s, the abundance
of $^{15}$N is not significantly reduced as described above, and thus
the results may not be changed substantially.

\subsubsection{Oxygen and fluorine}
\label{sec:of}

The abundance of oxygen is mainly correlated to $T_{\rm peak}$ but is
also dependent on $\tau$ (Figure~\ref{fig10}), owing to the presence
of three isotopes. The ratio $^{16}{\rm O}/^{17}{\rm O}$ has a clear
correlation to $T_{\rm peak}$. It reaches the minimum ($\sim 0.3$) at
$T_{\rm peak} \sim 3 \times 10^8$~K, and is nearly constant ($\sim 3$
for $T_{\rm peak} \lesssim 2 \times 10^8$~K and $\sim 10$ for $T_{\rm
peak} \gtrsim 4 \times 10^8$~K), due to the different nuclear reaction
cycles (Figure~\ref{fig5}). In contrast, the ratio $^{16}{\rm
O}/^{18}{\rm O}$ shows a clear correlation with the cooling timescale
(Figure~\ref{fig10}), being significantly small for $\tau \lesssim
100$~s. As a result, the abundance of oxygen reaches $\sim 0.03 - 0.1$
for (1) $T_{\rm peak} \lesssim 3 \times 10^8$~K ($^{16}$O and $^{17}$O
are abundant) or for (2) $\tau \lesssim 100$~s ($^{18}$O is
abundant). Note that oxygen is always abundant in the models $M_{\rm WD}
\lesssim 1.15 M_\odot$, where one of these conditions is satisfied.

Fluorine ($^{19}$F) is not significantly enhanced in the all models
(Figure~\ref{fig10}). The reason is that the reaction $^{18}$F($p$,
$\gamma$)$^{19}$Ne, which is followed by the $\beta^+$-decay to
$^{19}$F, is much slower than $^{18}$F($p$, $\alpha$)$^{15}$O. The
abundance of $^{19}$F is $\sim 10^{-4}$ at most for $\tau \sim 10$~s,
which is comparable to the $\beta^+$-decay lifetime of $^{19}$Ne
($\simeq 25$~s).

The oxygen-rich ONeMg novae ($\sim 0.1 - 0.3$ by mass) V693 CrA, QU Vul,
V351 Pup, and V1974 Cyg (Table~\ref{tab:obs}) can be explained by the
following models: (1) $M_{\rm WD} \lesssim 1.15 M_\odot$, (2) $T_{\rm
peak} \lesssim 2 \times 10^8$~K, or (3) $\tau \lesssim 10$~s. On the
other hand, V838 Her is fairly oxygen poor ($\simeq 3.3 \times
10^{-3}$), which could be explained by a rather massive model ($M_{\rm
WD} \sim 1.3 M_\odot$). It should be noted, however, that its estimated
metallicity is $\simeq 0.09$ (Table~\ref{tab:obs}), being significantly
less than assumed in this section (see \S~\ref{sec:depz}).

The ratio C/O can be $\gtrsim 1$ for $\tau \sim 1000$~s where the
abundance of carbon is $\sim 0.1$ and that of oxygen is $\lesssim
0.1$. It implies that the carbon-rich ONeMg novae, i.e., V1370 Aql and
V838 Her, may be explained by the models with $\tau \sim 1000$~s. Note
that carbon tends to be overproduced in the models with $\tau \gg
1000$~s (\S~\ref{sec:cn}). This may not, however, change the above
result with $\tau \sim 1000$~s.

\subsubsection{Neon and sodium}
\label{sec:nena}

Neon is the second most abundant metal in the initial composition
(Table~\ref{tab:init}). The abundance of neon is not significantly
reduced for $T_{\rm peak} \lesssim 4 \times 10^8$~K, due to its rather
slow proton capture (Figure~\ref{fig11}). Nevertheless, the
substantial nuclear flow appears in the Ne-Na cycle even for $T_{\rm
peak} \sim 2 - 3 \times 10^8$~K (Figure~\ref{fig5}) owing to the
abundant neon initially present. The ratio $^{20}{\rm Ne}/^{21}{\rm Ne}$
is clearly correlated with the cooling timescale, being small for the
shorter $\tau$, where the $\beta^+$-decay lifetime of $^{21}$Na ($\simeq
32$~s) is not negligible. On the other hand, the ratio $^{20}{\rm
Ne}/^{22}{\rm Ne}$ is clearly correlated to the peak temperature,
increasing with a rise in $T_{\rm peak}$. This is due to the faster
proton capture on $^{22}$Ne than on $^{20}$Ne.

The abundance of sodium is $\lesssim 10^{-3}$ for $\tau \gtrsim 100$~s,
due to the steady Ne-Na cycle where $^{20}$Ne is most abundant
(Figure~\ref{fig11}). The isotope ratio is also determined by their
reaction rates as $$ ^{22}{\rm Na}/^{23}{\rm Na} =
\lambda\left[^{23}{\rm Na}(p,\alpha)\right]/ \lambda\left[^{22}{\rm
Na}(p,\gamma)\right]\sim 10 \:, $$ in the temperature range $\sim 2 - 4
\times 10^8$~K. On the other hand, sodium is abundant ($\sim 0.01 - 0.1$
by mass) for $\tau \lesssim 100$~s, where the $\beta^+$-decay lifetimes
of $^{22}$Mg ($\simeq 6$~s) and $^{23}$Mg ($\simeq 16$~s) are not
negligible. Thus, a part of sodium, which is the decayed product of the
magnesium isotopes, survives the subsequent proton capture. The ratio
$^{22}{\rm Na}/^{23}{\rm Na}$ reaches $\sim 1$, owing to the abundant
$^{22}$Mg and $^{23}$Mg in the Ne-Na region during outbursts. The
abundance of $^{22}$Na shows a similar trend to that of sodium, clearly
correlated to the cooling timescale. This abundance can be changed by
the large uncertainty of the $^{22}$Na($p$, $\gamma$)$^{23}$Mg rate
(\cite{Kubo94}; \cite{Schm95}; \cite{Coc95}; \cite{Kubo96}). However, it
may not be significantly affected for $\tau \lesssim 100$~s, since the
explosive burning ceases while $^{22}$Mg is abundant.

The enrichment in neon is characteristic of all the observed ONeMg novae.
On the other hand, no positive detection of sodium has been reported for
recent ONeMg novae (\cite{Gehr94}), due to lack of useful lines and,
probably, little enrichment in sodium in the nova ejecta. An alternative
way to check the nucleosynthesis in the Ne-Na region is to compare with
the result of the $\gamma$-ray line survey of the $^{22}$Na decay from a
nearby ONeMg nova by CGRO or INTEGRAL in the near future.

\subsubsection{Magnesium and aluminum}
\label{sec:mgal}

Magnesium is one of the abundant elements initially present, but rather
fragile against proton capture. As a result, it is mostly transferred to
aluminum and silicon via the opened Mg-Al cycle (\cite{Timm88};
\cite{Cham88}). As seen in Figure~\ref{fig12}, the abundance of
magnesium reaches its minimum at $\tau \sim 1000$~s, in contrast to
carbon (Figure~\ref{fig9}). For $\tau \lesssim 1000$~s, it reaches
$\sim 10^{-2}$ due to the substantial leakage from the Ne-Na cycle and
the non-negligible $\beta^+$-decay lifetime of $^{25}$Al ($\simeq
10$~s). Note that the most abundant isotope is always $^{25}$Mg due to
its slowest proton capture. The isotope ratios $^{24}{\rm Mg}/^{25}{\rm
Mg}$ and $^{24}{\rm Mg}/^{26}{\rm Mg}$ are clearly correlated to the
cooling timescale. They are, however, not monotonic with $\tau$ but
complicated due to the inflow from the Ne-Na cycle and the leakage from
the Mg-Al region, and the various nuclear paths at high temperature
(Figure~\ref{fig5}).

The abundance of aluminum shows a similar trend to that of magnesium,
correlated to the cooling time- scale (Figure~\ref{fig12}). The ratio
$^{26}{\rm Al}/^{27}{\rm Al}$ is not significantly changed, being close
to $$^{26}{\rm Al}/^{27}{\rm Al} = \lambda\left[^{27}{\rm
Al}(p,\gamma)\right]/ \lambda\left[^{26}{\rm Al}(p,\gamma)\right]\sim
0.1-0.5 $$ in the temperature range $\sim 1 - 4 \times 10^8$~K. However,
the ratio decreases with a reduction in the cooling timescale, due to
the non-negligible $\beta^+$-decay lifetime of $^{27}$Si ($\simeq 6$~s)
which is the parent isotope of $^{27}$Al.  Note that, for rather high
temperature models ($T_{\rm peak} \gtrsim 4 \times 10^8$~K), the proton
capture on $^{25}$Al is faster than its $\beta^+$-decay. The subsequent
isotope $^{26}$Si decays to $^{26}$Mg in $\sim 12$~s through the
isomeric state of $^{26}$Al, bypassing its ground state. The double
peaks in $^{26}$Al ($\sim 3 \times 10^{-3}$ by mass) can be seen in
Figure~\ref{fig12}. The one at lower peak temperatures ($\sim 1.8
\times 10^8$~K) is consistent with PSTWS95, STWS98, and JH98, in which
the abundance of $^{26}$Al decreases with increasing white dwarf
mass. The other peak at higher peak temperatures ($\gtrsim 3 \times
10^8$~K) is the consequence of the substantial nuclear flux from the
Ne-Na region. The latter peak, which has not been presented in the
previous works, is of importance on whether ONeMg novae can be the
significant contributors of the Galactic $^{26}$Al. Note that the
abundance of $^{26}$Al in the latter case does not substantially depend
on the initial abundance of $^{24}$Mg (\S~\ref{sec:nzpl}). There are
large uncertainties in the reaction rates of $^{25}$Al($p$,
$\gamma$)$^{26}$Si (\cite{Wies86}; \cite{Coc95}; \cite{Ilia96}),
$^{26}$Si($p$, $\gamma$)$^{27}$P (\cite{Hern95}), $^{25}$Mg($p$,
$\gamma$)$^{26}$Al (\cite{Coc95}; \cite{Ilia96}), and $^{26}$Al($p$,
$\gamma$)$^{27}$Si (\cite{Coc95}; \cite{Cham93}; \cite{Coc95}). Our
trial calculations for a few models suggest that these uncertainties
change the abundance of $^{26}$Al by a factor of $\sim 2-3$.

The clear dependence of magnesium on the cooling timescale is useful to
constrain ($M_{\rm WD}$, $M_{\rm env}$) for observed ONeMg novae. The
estimated abundance of magnesium is $\sim 4 \times 10^{-3} - 2 \times
10^{-2}$ for V693 CrA, V1370 Aql, and QU Vul (Table~\ref{tab:obs}),
corresponding to $\tau \lesssim 100$~s or $\tau \gtrsim 10^6$~s (see
Figures~\ref{fig6} and \ref{fig12}). The abundance of
aluminum does not significantly vary in the $M_{\rm WD}$--$M_{\rm env}$
space, being not useful to constrain ($M_{\rm WD}$, $M_{\rm
env}$). Nevertheless, the abundance estimates of aluminum are $\sim 3
\times 10^{-3} - 10^{-2}$ for V693 CrA, QU Vul, and V351 Pup
(Table~\ref{tab:obs}), which is in good agreement with our results.

\subsubsection{Silicon and phosphorus}
\label{sec:sip}

The abundance of silicon reaches $\sim 3 \times 10^{-2}$ for $T_{\rm
peak} \gtrsim 2 \times 10^8$~K (Figure~\ref{fig13}) via the
substantial nuclear flux from the Mg-Al region. The abundance is only
weakly correlated to the cooling timescale. On the other hand, the
ratios $^{28}{\rm Si}/^{29}{\rm Si}$ and $^{28}{\rm Si}/^{30}{\rm Si}$
are clearly correlated to the cooling timescale, because of various
competitions between proton capture and $\beta^+$-decay
(Figure~\ref{fig5}).

The abundance of phosphorus ($^{31}$P) reaches $\sim 10^{-3} - 10^{-2}$
for $T_{\rm peak} \gtrsim 3 \times 10^8$~K, due to the faster proton
capture on $^{30}$P than its $\beta^+$-decay
(Figure~\ref{fig13}). Since the Si-P cycle is not closed as seen in
Figure~\ref{fig5}, phosphorus is not significantly destroyed.

The abundance of silicon in the ejecta of V693 CrA, V1370 Aql, and V351
Pup is as small as $\sim 2 - 7 \times 10^{-3}$, corresponding to $T_{\rm
peak} \lesssim 2 \times 10^8$~K. In contrast, that in QU Vul ($\sim 3 -
4 \times 10^{-2}$) is in agreement with the models $T_{\rm peak} \gtrsim
2 \times 10^8$~K. The discovery of phosphorus has been reported in the
ejected shell of V1974 Cyg by a near infrared spectroscopy
(\cite{Wagn96}). It suggests that V1974 Cyg can be explained by the
model with a rather high peak temperature, although an accurate
abundance of phosphorus is required to constrain ($M_{\rm WD}$, $M_{\rm
env}$). It is also interesting to note that significantly enhanced
phosphorus has been detected on the white dwarf in a dwarf nova system
(\cite{Sion97}) and in the broad line system of a QSO (\cite{Shie96}),
which might originate from ONeMg novae.

\subsubsection{Sulfur and other heavy elements}
\label{sec:sca}

The abundance of sulfur reaches $\sim 10^{-2}$ for $T_{\rm peak} \gtrsim
3 \times 10^8$~K, through leakage from the Si-P region
(Figure~\ref{fig14}). The abundance does not exceed $10^{-2}$ in the
models $M_{\rm WD} \lesssim 1.15 M_\odot$ because of the shorter cooling
timescale (Figure~\ref{fig6}). This condition is, however, highly
dependent on the initial composition as will be discussed in
\S~\ref{sec:depz}. For $T_{\rm peak} \gtrsim 3 \times 10^8$~K, the
ratios $^{32}{\rm S}/^{33}{\rm S}$ and $^{32}{\rm S}/^{34}{\rm S}$
decrease with a rise in peak temperature, due to the increasing nuclear
paths (Figure~\ref{fig5}). For $T_{\rm peak} \lesssim 3 \times
10^8$~K, these ratios approach those determined by their reaction rates.

At least a half of the observed ONeMg novae, V1370 Aql, QU Vul, and V838
Her, are abundant in sulfur in their ejecta (Table~\ref{tab:obs}). In
addition, the sulfur enrichment has been confirmed in the V1974 Cyg
ejecta from near infrared spectroscopies (\cite{Wood92}; \cite{Wood95};
\cite{Wagn96}). These novae can be explained by the models with such
high peak temperatures as $T_{\rm peak} \gtrsim 3 \times 10^8$~K. The
estimated abundance of sulfur for V1370 Aql is much more abundant than
by any models in the $M_{\rm WD}$--$M_{\rm env}$ space
(Figure~\ref{fig14}). It should be noted, however, the estimated
metallicity for V1370 Aql is twice as much as assumed in this section
(see \S~\ref{sec:depz}).

Heavier elements, from chlorine to calcium, are not substantially
enhanced for $T_{\rm peak} \lesssim 4 \times 10^8$~K
(Figure~\ref{fig14}). In addition, their enhancement is never seen
in the models $M_{\rm WD} \lesssim 1.15 M_\odot$ due to the shorter
cooling timescale. Nevertheless, the enrichment in chlorine has been
reported for the ejecta of V1974 Cyg by a near infrared spectroscopy
(\cite{Wagn96}). The accurate abundance of chlorine would severely
constrain ($M_{\rm WD}$, $M_{\rm env}$) for V1974 Cyg.

\subsection{Dependence on the initial composition}
\label{sec:depz}

So far we have discussed the nucleosynthesis results for only one set of
the initial composition $X_{\rm WD} = 0.4$ (case~B). However, the
metallicities of the ejecta for V1370 Aql, QU Vul by Saizar et
al. (1992), and V838 Her are significantly deviated from 0.4
(Table~\ref{tab:obs}). In addition, the different authors present
different metallicity estimates for the same nova events. In particular,
the discrepancy is serious for QU Vul between Saizar et al. (1992)
($\simeq 0.10$) and Andre\"a et al. (1994) ($\simeq 0.40$). It is,
therefore, difficult to judge whether the dispersion of the
metallicities is real or due to observational errors. In the following,
we discuss how the initial composition influences the nucleosynthesis
results, comparing the low ($X_{\rm WD} = 0.1$; case~A) and high
($X_{\rm WD} = 0.8$; case~C) metallicity cases.

As discussed in \S~\ref{sec:model}, the density and temperature
structures of an envelope are determined uniquely by a set of ($M_{\rm
WD}$, $M_{\rm env}$) in our model, being independent of its time
evolution (but slightly dependent on the time variation in mean
molecular weight). As a result, case~C is at most 20~\% higher than
case~A in peak temperature for each ($M_{\rm WD}$, $M_{\rm env}$) as
seen in Figure~\ref{fig15}. The higher temperature in case~C is due
to the larger mean molecular weight. In contrast, a variation in initial
composition is crucial for the cooling timescale
(Figure~\ref{fig15}). For $T_{\rm peak} \gtrsim 2 \times 10^8$~K,
case~C is more than 10 times shorter than case~A in $\tau$. This is a
consequence of the higher nuclear energy in case~C
(Figure~\ref{fig16}) due to the abundant nuclear fuel. The ejection
velocity is also affected by the initial composition. As seen in
Figure~\ref{fig16}, case~C obtains significantly higher $v_{\rm
ej}$ than case~A in each model.

A prominent distinction between case~A (N0540A) and case~C (N0540C) can
be seen in Figure~\ref{fig17}, which shows the nuclear flows and
the final yields in the model ($M_{\rm WD}/M_\odot$, $M_{\rm
env}/M_\odot$) = (1.05, 10$^{-4.0}$). In N0540A, the nuclear flow
extends to sulfur due to the longer cooling timescale ($\simeq
23000$~s), while that in N0540C ($\tau \simeq 1800$~s) to silicon. The
model N0540A consumes most of oxygen initially present, in contrast to
N0540C.

Figures~\ref{fig18}--\ref{fig20} show the abundances of
important elements and $\gamma$-ray emitters in the
$M_{\rm WD}$--$M_{\rm env}$ space for case~A and case~C.
These results are explained as follows:\\
1. The abundance of carbon is still clearly correlated to $\tau$ as
in case~B (\S~\ref{sec:cn}), reaching its maximum at $\tau \sim 1000$~s
for both cases (Figure~\ref{fig18}). The abundance is roughly
proportional to $X_{\rm WD}$ among the models with the same cooling
timescale.\\
2. Magnesium is another element clearly correlated to $\tau$ as in case~B
(\S~\ref{sec:mgal}). In case~A, the abundance is significantly smaller
than in case~C, not enhanced even for $\tau \lesssim 1000$~s. This is a
consequence of the longer $\tau$ in case~A, where the nuclear flow
extends to heavier elements than magnesium (Figure~\ref{fig17}).\\
3. Silicon is also an element showing a correlation to $\tau$ in case~B,
not significantly changed for $T_{\rm peak} \gtrsim 2 \times 10^8$~K
(\S~\ref{sec:sip}). This feature holds for case~C. However, the
abundance in case~A has a correlation to $T_{\rm peak}$
rather than $\tau$, reaching its maximum at
$T_{\rm peak} \sim 2.5 \times 10^8$~K (Figure~\ref{fig19}).
The depletion of silicon in case~A for high $T_{\rm peak}$ is due to
the long cooling timescale.\\
4. The trend of oxygen abundance significantly differs between case~A
and case~C (Figure~\ref{fig18}). The abundance
in case~B is correlated to both $T_{\rm peak}$ and $\tau$, being more
abundant in the lower $M_{\rm WD}$ models (\S~\ref{sec:of}).
In case~C, however, the abundance is not significantly changed in the
($M_{\rm WD}$, $M_{\rm env}$) space, being $\sim 0.3$.
On the other hand, that in case~A is clearly correlated to
the peak temperature, significantly depleted for $T_{\rm peak}
\gtrsim 2.5 \times 10^8$~K (Figure~\ref{fig18}).\\
5. The abundance of sulfur shows a correlation to $T_{\rm peak}$ in
all cases. In case~C, however, the abundance is $\lesssim
10^{-3}$ for the models $M_{\rm WD} \lesssim 1.15 M_\odot$ because of
the shorter $\tau$. On the other hand, that in case~A
reaches $\sim 3\times 10^{-2}$ at $\gtrsim 3 \times 10^8$~K, since
$\tau$ is longer and thus the nuclear flow extends to heavier elements.\\
6. The radioactive species $^7$Be, $^{22}$Na, and $^{26}$Al are not
significantly enhanced in case~A because of its longer cooling timescale
(Figure~\ref{fig20}). On the other hand, these abundances in
case~C show similar trends to case~B in the $M_{\rm WD}$--$M_{\rm env}$
space (Figures~\ref{fig8}).

The estimated metallicity for the ejecta of V1370 Aql is extremely high
($Z \sim 0.85$; Table~\ref{tab:obs}), which is close to case~C. However,
the abundance of oxygen is significantly small ($\sim 4 - 5 \times
10^{-2}$), being inconsistent with our results ($\gtrsim 0.1$). In
addition, sulfur in the ejecta is extremely abundant ($\sim 0.1$ by
mass), which is also in disagreement with our results ($\lesssim
10^{-2}$). These features, i.e., the abundances of low oxygen and high
sulfur, could be explained by lower $X_{\rm WD}$ models rather than
higher ones (Figure~\ref{fig10}, \ref{fig14}, \ref{fig18},
and \ref{fig19}). Thus, this extremely high metallicity in this
nova ejecta may not be real but due to the difficulties in the
observational estimations.

\begin{table*}[t]
\caption{Ejected Masses of recent ONeMg Novae}
\label{tab:menv}
\smallskip
\begin{tabular}{lrl}
\hline
\hline
  & $M_{\rm ej}/M_\odot$ & Observations \\
\hline
QU Vul\tablenotemark{1}    & $8\times 10^{-4}$       & Radio emission \\
QU Vul\tablenotemark{2}    & $\ge 9\times 10^{-4}$   & Infrared emission \\
QU Vul\tablenotemark{3}    & $0.2-1.5\times 10^{-4}$ & Multiwavelength study \\
V351 Pup\tablenotemark{4}  & $1\times 10^{-7}$       & Multiwavelength study \\
V838 Her\tablenotemark{5}  & $6.4-9\times 10^{-5}$   & Infrared emission \\
V838 Her\tablenotemark{6}  & $1.8\times 10^{-4}$     & Optical and UV emission \\
V1974 Cyg\tablenotemark{7} & $\ge 7\times 10^{-5}$   & Radio emission \\
V1974 Cyg\tablenotemark{8} & $1-4\times 10^{-4}\times Y^{-1/2\;\dag}$
                                                     & UV emission \\
V1974 Cyg\tablenotemark{9} & $2-5\times 10^{-4}$     & Infrared emission \\
\hline
\end{tabular}

\smallskip
{\footnotesize
References: $^1$Taylor et al. 1987, $^2$Greenhouse et al. 1988, 
$^3$Saizar et al. 1992, $^4$Saizar et al. 1996, 
$^5$Woodward et al. 1992, $^6$Vanlandingham et al. 1996, 
$^7$Pavelin et al. 1993, $^8$Shore et al. 1993, $^9$Woodward et al. 1997

$^{\dag}Y$ is the enhancement factor for the helium abundance
}
\end{table*}

For the QU Vul ejecta, Saizar et al. (1992) gave a much lower
metallicity estimate ($Z \simeq 0.10$) corresponding to case~A than
Andre\"a et al. (1994). The low abundance estimates of carbon, oxygen,
and magnesium by Saizar et al. (1992) are in good agreement with our
results for $T_{\rm peak} \lesssim 2 \times 10^8$~K
(Figures~\ref{fig15}, \ref{fig18}, and
\ref{fig19}). However, the abundance of silicon ($\sim 4 \times
10^{-2}$ by mass) suggests that the nova has obtained $T_{\rm peak} \sim
2 - 3 \times 10^8$~K, which is inconsistent with the above result. Thus,
there is no ($M_{\rm WD}$, $M_{\rm env}$) model which explains the
abundance estimates by Saizar et al. (1992) within reasonable
observational errors.

The V838 Her ejecta also shows a rather low metallicity estimate ($Z
\simeq 0.09$), which again corresponds to case~A. The abundance features
of the ejected shell, i.e., the low oxygen and high sulfur, are well
reproduced in our results for $T_{\rm peak} \sim 2 - 3 \times 10^8$~K
(Figures~\ref{fig18} and \ref{fig19}). Hence, the low
metallicity for this case implies the presence of a real dispersion in
metallicity among the observed nova ejecta.

\section{Comparison with observations}
\label{sec:obs}

In this section, we discuss which ($M_{\rm WD}$, $M_{\rm env}$) models
best match the recent ONeMg nova observations from the nucleosynthetic
point of view, using the results of case~B ($X_{\rm WD} = 0.4$). For
V838 Her, however, those of case~A ($X_{\rm WD} = 0.1$) are used
(\S~\ref{sec:depz}). The abundances for QU Vul by Saizar et al. (1992)
and V1370 Aql are not discussed in this section, since they are not
reproduced in our models (\S~\ref{sec:depz}).

Figure~\ref{fig21} shows the models which are in agreement with the
abundance estimates for recent ONeMg novae, within a factor of three for
V693 CrA (\cite{Vanl97}; triangles), V351 Pup (\cite{Saiz96};
asterisks), and V1974 Cyg (\cite{Aust96}; stars), and of five for QU Vul
(\cite{Andr94}; circles) and V838 Her (\cite{Vanl97}; squares). The
thick symbol for each nova is the best model, whose ratio to its
observation is shown in Figure~\ref{fig22}. Interestingly, at least
four events (V693 CrA, QU Vul, V838 Her, and V1974 Cyg) are well
explained by the models $\simeq 1.1 M_\odot$, which is near the lower
limit to ONeMg cores (\cite{Nomo84}). This is in contrast to the mass
range of $1.25 - 1.35 M_\odot$ used by PSTWS95 and STWS98, that is near
the upper bound to ONeMg cores. Table~\ref{tab:menv} shows the estimated
ejecta masses of QU Vul (\cite{Tayl87}; \cite{Gree88}; \cite{Saiz92}),
V351 Pup (\cite{Saiz96}), V838 Her (\cite{Wood92}; \cite{Vanl96}), and
V1974 Cyg (\cite{Pave93}; \cite{Shor93}; \cite{Wood97}) from
observations. These significantly high ejecta masses compared with
theoretical estimates are reasonably explained by our nucleosynthesis
results if we assume that almost all the envelope is eventually blown
off. In addition, for the models with $M_{\rm env} \gtrsim 10^{-4}
M_\odot$, the expansion velocities exceed $v_{\rm esc}$ and obtain
$v_{\rm ej} \gtrsim 1000$~km s$^{-1}$ (Figures~\ref{fig6} and
\ref{fig16}), which are in good agreement with observations. Note
that the abundances of carbon and nitrogen by our results are also in
good agreement with those by observations, regardless of their
uncertainties (\S~\ref{sec:comp}). This is a consequence that these
novae are well explained by the models with $\tau \lesssim 1000$~s where
the uncertainties (caused by the depletion of $^{15}$N) may be small
(\S~\ref{sec:cn}).

\subsection{V693 CrA}
\label{sec:v693}

The high oxygen abundance ($\sim 0.1 - 0.2$ by mass) in the V693 CrA
ejecta (\cite{Will85}; \cite{Andr94}; \cite{Vanl97}) implies that it was
an event with $T_{\rm peak} \lesssim 2 \times 10^8$~K or with $M_{\rm
WD} \lesssim 1.15 M_\odot$ (\S~\ref{sec:of}). The low magnesium and high
silicon abundances by Vanlandingham, Starrfield, \& Shore (1997) suggest
that the cooling timescale was $\lesssim 1000$~s (\S~\ref{sec:mgal} and
\ref{sec:sip}). On the other hand, Williams et al. (1985) and Andre\"a,
Drechsel, \& Starrfield (1994) present somewhat higher magnesium and
lower silicon abundances. We compare our results with the abundance
estimates by Vanlandingham, Starrfield, \& Shore (1997), since others
used the overexposed spectrum as pointed out by Andre\"a, Drechsel, \&
Starrfield (1994). As a result, the model ($M_{\rm WD}/M_\odot$, $M_{\rm
env}/M_\odot$) = (1.05, $10^{-3}$) (case~B) is in good agreement with
the observation within a factor of 3 (Figures~\ref{fig21} and
\ref{fig22}).

\subsection{QU Vul}
\label{sec:qu}

The high abundance of sulfur implies that the nova obtained such a high
temperature as $T_{\rm peak} \gtrsim 3 \times 10^8$~K
(Figures~\ref{fig6} and \ref{fig14}). Furthermore, the
abundance of oxygen despite such a high temperature suggests that the
white dwarf mass was $\lesssim 1.15 M_\odot$ (\S~\ref{sec:of}). Our
results are in agreement with the observational estimates within a
factor of 5 for the models ($M_{\rm WD}/M_\odot$, $M_{\rm env}/M_\odot$)
= ($1.05 - 1.1$, $10^{-3.5} - 10^{-3}$) (case~B). These high envelope
masses are in good agreement with the observational estimates of the
nova ejecta (Table~\ref{tab:menv}). Note that the both high abundances
of oxygen and sulfur were not explained by previous hydrodynamic
studies, with much smaller envelope masses.

\subsection{V351 Pup}
\label{sec:v351}

The ejected shell of V351 Pup shows the high oxygen and low silicon
abundances (\cite{Saiz96}). This feature is well explained with the low
temperature models of $T_{\rm peak} \lesssim 2 \times 10^8$~K
(Figures~\ref{fig6}, \ref{fig10}, and \ref{fig13}). Our
results are in good agreement with the observational estimates within a
factor of 3 for the models ($M_{\rm WD}/M_\odot$, $M_{\rm env}/M_\odot$)
= ($1.05 - 1.1$, $10^{-5.5} - 10^{-5}$), ($1.15 - 1.2$, $10^{-6} -
10^{-5.5}$), and the best for (1.25, $10^{-6}$) (case~B). In such low
temperature models, magnesium must be abundant (\S~\ref{sec:mgal}),
though it is not presented in Saizar et al. (1996). The above low
envelope masses may be due to mass accreting at high rate from a giant
companion, that is also suggested by the optical spectral analysis
(\cite{Saiz96}). The estimated ejecta mass, $2 \times 10^{-7} M_\odot$
(Table~\ref{tab:menv}), implies that this nova occurred in such a
massive white dwarf as $M_{\rm WD} \gtrsim 1.25 M_\odot$.

\subsection{V838 Her}
\label{sec:v838}

The low oxygen and high sulfur abundances in the V838 Her ejecta are the
prominent feature in the low metallicity models (case~A) with $T_{\rm
peak} \sim 2.5 - 3 \times 10^8$~K (\S~\ref{sec:depz}). In addition, the
ratios ${\rm C}/{\rm N} \sim 1$ and ${\rm C}/{\rm O} \gtrsim 1$ suggest
that the cooling timescale was $\sim 1000$~s (\S~\ref{sec:cn} and
\ref{sec:of}). Thus, the nova may have occurred with the low $M_{\rm
WD}$ and the high $M_{\rm env}$ (Figure~\ref{fig15}). Our results
are in agreement with the observational estimates within a factor of 5
for the model ($M_{\rm WD}/M_\odot$, $M_{\rm env}/M_\odot$) = (1.05,
$10^{-4} - 10^{-3.5}$).

\subsection{V1974 Cyg}
\label{sec:v1974}

Unfortunately, the abundances heavier than neon are not presented in
Austin et al. (1996), due to lack of these lines.  The high oxygen
abundance suggests that the peak temperature was $\lesssim 2 \times
10^8$~K or the white dwarf mass was $\lesssim 1.15 M_\odot$
(\S~\ref{sec:of}). In addition, the ratio ${\rm C}/{\rm N} \sim 1$
implies that the cooling timescale was $\sim 1000$~s
(\S~\ref{sec:cn}). Our results are in good agreement with the
observational estimates within a factor of 3 for the models ($M_{\rm
WD}/M_\odot$, $M_{\rm env}/M_\odot$) = (1.05, $10^{-3.5}$), (1.1,
$10^{-4}$), and (1.2, $10^{-5}$), and the best for (1.1, $10^{-4.5}$)
(case~B; N0535B, N1040B, N2050B, and N1045B). They are in reasonable
agreement with the estimated mass of the ejecta $\gtrsim 5 \times
10^{-5} M_\odot$ (Table~\ref{tab:menv}). Their white dwarf masses are
also in agreement with the estimates from observations $\sim 0.75 - 1.1
M_\odot$ (\cite{Pare95}; \cite{Rett97}) but smaller than $\sim 1.25
M_\odot$ (\cite{Krau96}). The observation shows a factor of 2 lower
hydrogen abundance than our result (Figure~\ref{fig22}). This might
be due to the subsequent steady hydrogen burning on the white dwarf as
pointed out by Krauter et al. (1996). Hayward et al. (1996) have derived
the neon and magnesium abundances relative to solar values from a
mid-infrared observation. If their ratio ${\rm Ne}/{\rm Mg} \sim 30$ is
adopted, the abundance of magnesium would be $\sim 3 \times 10^{-3}$. It
favors a relatively high envelope mass model
(Figure~\ref{fig12}). As a result, N1040B would be the best in this
case. A recent near infrared measurement has shown the presence of the
lines of phosphorus and chlorine together with sulfur in the V1974 Cyg
ejecta (\cite{Wagn96}). This suggests that V1974 Cyg experienced $T_{\rm
peak} \gtrsim 3 \times 10^8$~K. In this case, the higher $M_{\rm env}$
models are also favorable. In addition, the ejection velocity in N1040B
is $\simeq 1800$~km s$^{-1}$ being good agreement with observations
($\simeq 2300$~km s$^{-1}$; \cite{Gehr98}), while that in N1045B is
$\simeq 190$~km s$^{-1}$. Obviously, further analysis of heavy elements
is needed to constrain the parameters ($M_{\rm WD}$, $M_{\rm env}$) for
V1974 Cyg.

\section{Production of the radioactive isotopes}
\label{sec:gamma}

\begin{table*}[t]
\caption{$^{22}$Na Production in ONeMg Novae}
\label{tab:na22}
\smallskip
\begin{tabular}{cccccc}
\hline
\hline
 & year & $d$ (kpc) & $XM_{\rm env}/M_\odot$
 & $F_0$\tablenotemark{a} (cm$^{-2}$ s$^{-1}$)
 & $F_{\rm up}$\tablenotemark{b} (cm$^{-2}$ s$^{-1}$) \\
\hline
V693 CrA  & 1981 & 11.8 & $2.0\times 10^{-5}$  & $5.5\times 10^{-4}$ &                    \\
QU Vul    & 1984 &  2.8 & $8.9\times 10^{-7}$  & $4.4\times 10^{-4}$ &                    \\
V351 Pup  & 1991 &  3.5 & $4.0\times 10^{-11}$ & $1.3\times 10^{-8}$ & $5.5\times 10^{-5}$ \\
V838 Her  & 1991 &  3.4 & $1.9\times 10^{-8}$  & $6.3\times 10^{-6}$ & $3.3\times 10^{-5}$ \\
V1974 Cyg & 1992 &  1.8 & $2.5\times 10^{-8}$  & $3.0\times 10^{-5}$ & $2.3\times 10^{-5}$ \\
\hline
\end{tabular}

\smallskip
{\footnotesize
$^a$initial flux by this work (best sequence)

$^b$upper limit by Iyudin et al. (1995)
}
\end{table*}

In this section, we discuss the possibilities of detecting the
$\gamma$-ray emitters $^7$Be, $^{22}$Na, and the contribution to the
Galactic $^{26}$Al, based on our nucleosynthesis results in ONeMg novae.
Figure~\ref{fig23} shows the total masses of $^7$Be, $^{22}$Na,
and $^{26}$Al produced per event for $X_{\rm WD} = 0.4$ (case~B). As
seen in this figure, the models $M_{\rm WD} \simeq 1.1 M_\odot$ with
$M_{\rm env} \gtrsim 10^{-4} M_\odot$, which are in good agreement with
most of observations (\S~\ref{sec:obs}), produce significant amounts of
these isotopes. In the rest of this section, all the envelope is assumed
to be eventually blown off.

\subsection{Gamma-ray emission from $^7$Be electron captures}
\label{sec:be7}

Classical novae might be a possible site for the $^7$Li production (the
electron-capture products of $^7$Be) in the solar neighborhood
(\cite{Star78}; \cite{Dant91}; \cite{Wana98}). Recently, some works
claim that they cannot be the major contributors (\cite{Matt95};
JH98). Nevertheless, the $\gamma$-rays (at 478~KeV) from the $^7$Be
electron capture would be detectable by CGRO or INTEGRAL (\cite{Hern96};
\cite{Jose98}; \cite{Gome98}).

The $\gamma$-ray line flux from the $^7$Be electron capture is estimated
as
\begin{eqnarray}
F\left(^7{\rm Be}\right)
& \sim & 4\times 10^{-5} {\rm cm}^{-2} {\rm s}^{-1} \nonumber \\
& \times &
      \left(\frac{X\left(^7{\rm Be}\right)M_{\rm env}}
      {5\times 10^{-9}M_\odot}\right) 
      \left(\frac d{\rm 3~kpc}\right)^{-2} \nonumber \\
& \times & e^ {-t/\tau \left(^7{\rm Be}\right)} \:, \nonumber
\end{eqnarray}
where $d$ is the distance of the nova system from the sun. For the CGRO
sensitivity (a few $10^{-5}$~cm$^{^2}$~s$^{-1}$) and a typical distance
($d \sim 3$~kpc), the mass of $^7$Be $\sim 5 \times 10^{-9} M_\odot$ per
event is required to be detected. The mass of $^7$Be per event in our
model is over 5--10 times smaller than required
(Figure~\ref{fig23}). However, an ONeMg nova will be a promising
target of $^7$Be $\gamma$-rays for INTEGRAL in the near future. Note
that CO novae may produce about 10 times higher $^7$Be than ONeMg novae
(JH98; \cite{Wana98}).

\subsection{Gamma-ray emission from $^{22}$Na decays}
\label{sec:na22}

There has been increasing expectations that an ONeMg nova might be the
first stellar object for the detection of the $\gamma$-ray emitter
$^{22}$Na (\cite{Weis90}; \cite{Star93}; \cite{Coc95}; PSTWS95; Wanajo
1997a, b; STWS98; JH98; \cite{Gome98}). Nevertheless, no positive
detection has been reported by COMPTEL on board CGRO for the recent
ONeMg novae, V351 Pup, V838 Her, and V1974 Cyg (\cite{Iyud95}).

As seen in Figure~\ref{fig23}, the total mass of $^{22}$Na per
event (case~B) is significantly high in the models $M_{\rm env} \gtrsim
10^{-4}M_\odot$. The $\gamma$-ray line flux from the $\beta^+$-decay of
$^{22}$Na is estimated as
\begin{eqnarray}
F\left(^{22}{\rm Na}\right)
& \sim & 4\times 10^{-5} {\rm cm}^{-2} {\rm s}^{-1} \nonumber \\
& \times &
  \left(\frac{X\left(^{22}{\rm Na}\right)M_{\rm env}}
             {1\times 10^{-7}M_\odot}\right) 
  \left(\frac d{\rm 3~kpc}\right)^{-2} \nonumber \\
& \times & e^{-t/\tau \left(^{22}{\rm Na}\right)} \:. \nonumber
\end{eqnarray}
For the CGRO sensitivity and a typical distance ($\sim 3$~kpc), the mass
of $^{22}$Na $\sim 10^{-7} M_\odot$ per event is required. This
corresponds to an envelope mass of $\sim 10^{-4} M_\odot$
(Figure~\ref{fig23}). Table~\ref{tab:na22} shows the distances,
the masses of $^{22}$Na per event from our results (for the best fitted
models, see \S~\ref{sec:obs}), the expected initial $\gamma$-ray line
fluxes at 1.275~MeV, and the upper limits to the COMPTEL observations
(\cite{Iyud95}) for V693 CrA, QU Vul, V351 Pup, V838 Her, and V1974 Cyg. 
V1370 Aql is omitted here, not explained by any ($M_{\rm WD}$, $M_{\rm
env}$) models as discussed in \S~\ref{sec:obs}. V693 CrA and QU Vul may
have emitted the $\gamma$-rays as high as $\sim 5 \times
10^{-4}$~cm$^{-2}$s$^{-1}$. However, their fluxes have decreased to,
respectively, $6 \times 10^{-6}$ and $1 \times
10^{-5}$~cm$^{^2}$~s$^{-1}$ at present, and will decrease to $3 \times
10^{-6}$ and $5 \times 10^{-6}$~cm$^{^2}$~s$^{-1}$ at the launch of
INTEGRAL ($\sim 2001?$). In contrast to the above two novae, V351 Pup
and V838 Her may have yielded much lower $\gamma$-ray fluxes, which are
consistent with the upper limits by COMPTEL. The envelope mass of V351
Pup might be $\lesssim 10^{-5}M_\odot$ (\S~\ref{sec:v351}) so that
little $^{22}$Na may have been produced. Although V838 Her may have
obtained a massive envelope such as $\sim 10^{-4}M_\odot$
(\S~\ref{sec:v838}), the cooling timescale was so long owing to the low
metallicity ($\sim 10^4$~s) that little $^{22}$Na survived. The
$\gamma$-ray flux of the $^{22}$Na decay from V1974 Cyg may have been
near the sensitivity limit to COMPTEL, with the abundance in the best
model (\S~\ref{sec:v1974}) and the distance of $\sim 1.8$~kpc
(\cite{Choc97}). Thus, if the ejected mass was as high as a few $10^{-4}
M_\odot$ indeed, our model would have produced observable $^{22}$Na (or
the estimated distance is too short).

It seems that at least four ONeMg novae (V693 CrA, QU Vul, V838 Her, and
V1974 Cyg) in the past twenty years have produced sufficient $^{22}$Na
for the high sensitivity of INTEGRAL ($\sim 4 - 5 \times
10^{-6}$~cm$^{^2}$~s$^{-1}$). The next ONeMg nova in the first decade of
the 21st century will be a promising candidate for detecting the
$\gamma$-ray emitter $^{22}$Na.

\subsection{Galactic $^{26}$Al production}
\label{sec:al26}

Since the discovery by HEAO 3 (Mahoney et al. 1984), many studies have
been carried out to explain the presence of $\sim 1 - 3 M_\odot$ of
$^{26}$Al in the Galaxy. In particular, ONeMg novae have been considered
to be a promising stellar site for the $^{26}$Al production
(\cite{Weis90}; \cite{Nofa91}; \cite{Star93}; \cite{Coc95}; PSTWS95;
\cite{Kolb97}; \cite{Jose97}; STWS98; JH98), as well as AGB stars
(\cite{Fore91}), Wolf-Rayet stars (\cite{Pran86}; \cite{Meyn97}), and
Type II supernovae (\cite{Walt89}; \cite{Pran93};
\cite{Timm95}). However, the detailed observations of $\gamma$-ray lines
at 1.8~MeV by COMPTEL (\cite{Dieh94}; \cite{Dieh95}) have shown that the
Galactic $^{26}$Al originates from the youngest stellar population
associated with the spiral arms and the local groups (\cite{Pran96a};
\cite{Pran96b}). This may imply that the major sources of the Galactic
$^{26}$Al are Type II supernovae or Wolf-Rayet stars.

The mass of $^{26}$Al per event is up to $\sim 3 \times 10^{-7}M_\odot$
in the models $M_{\rm env} \gtrsim 10^{-4}M_\odot$
(Figure~\ref{fig23}). The upper limit to the Galactic $^{26}$Al
from ONeMg novae is thus estimated as
{\small
\begin{eqnarray}
M\left(^{26}{\rm Al}\right) & \sim & 3 M_\odot \nonumber \\
 & \times &
 \left(\frac{R_{\rm nova}}{40 {\rm yr}^{-1}}\right)
 \left(\frac{f_{\rm ONeMg}}{0.25}\right)
 \left(\frac{X(^{26}{\rm Al})M_{\rm env}}{3\times 10^{-7}}\right) \:, \nonumber
\end{eqnarray}
}
where $R_{\rm nova}$ is the nova rate in the Galaxy and $f_{\rm ONeMg}$
is the fraction of ONeMg novae. This is in good agreement with the
estimate from the CGRO results. If ONeMg novae are not be the major
contributors to the Galactic $^{26}$Al, its typical mass per event must
be somewhat smaller than the above value. There are some uncertainties
in the Galactic nova rates (\cite{Yung97}; \cite{Shaf97}; \cite{Hata97})
and the fraction of ONeMg novae (\cite{Ritt91}; \cite{Livi94}). However,
these uncertainties may be much smaller (a factor of $\sim 2$) than
those in the $^{26}$Al yields (about 2 orders of magnitude as can be
seen in Figure~\ref{fig23}). The INTEGRAL survey on the diffuse
component of the Galactic $^{26}$Al, together with a search of $^{22}$Na
from an individual ONeMg nova, will impose a severe constraint on the
current nova models.

\section{Conclusions}
\label{sec:concl}

In this paper we have examined nucleosynthesis in ONeMg novae with the
wide ranges of the three parameters, i.e., the white dwarf mass ($M_{\rm
WD} = 1.05 - 1.35 M_\odot$), the envelope mass ($M_{\rm env} = 10^{-6} -
10^{-3} M_\odot$), and the initial metallicity ($X_{\rm WD} = 0.1 -
0.8$). We used a quasi-analytic nova model with a one-zone envelope,
coupled with an updated nuclear reaction network code. Our
nucleosynthesis results are in good agreement with those of previous
hydrodynamic calculations except for several fragile isotopes.

We have found that the explosion is more violent in a lower $M_{\rm WD}$
model among those with the same peak temperature, due to its smaller
gravitational potential. There exists the critical cooling timescale
($\sim 1000$~s), at which the energy generation by the $\beta^+$-decay
of $^{14}$O and $^{15}$O plays a crucial role to the envelope expansion.
For the models with $\tau \lesssim 1000$~s, the nucleosynthesis results
significantly deviate from those expected in steady nuclear flows (e.g.,
the CNO and Ne-Na cycle). These models also obtain high ejection
velocities ($\gtrsim 1000$~km~s$^{-1}$), which are in good agreement
with recent observations.

There are a couple of characteristic trends for the abundances in
the $M_{\rm WD}$--$M_{\rm env}$ space as follows (case~B):\\
1. The abundances of oxygen, neon, phosphorus, and sulfur (and $^7$Be)
are clearly correlated to the peak temperatures, although those of
oxygen and sulfur are also dependent of the cooling timescales.
The abundance of oxygen is always abundant in the models
$M_{\rm WD} \lesssim 1.15 M_\odot$. The heavier elements than sulfur
show no significant enrichment in the models with
$T_{\rm peak} \lesssim 4 \times 10^8$~K.\\
2. The abundances of carbon, fluorine, sodium, and magnesium
(and $^{22}$Na, $^{26}$Al) are clearly correlated to the cooling
timescales. The abundance of $^{22}$Na is significantly high in the
models with $\tau \lesssim 100$~s. On the other hand, that of $^{26}$Al
shows double peaks in the $M_{\rm WD}$--$M_{\rm env}$ space.\\
3. The abundances of nitrogen, aluminum, and silicon are not
significantly changed in the $M_{\rm WD}$--$M_{\rm env}$ space,
although those are weakly dependent of the cooling timescales.

The initial metallicity $X_{\rm WD}$, as well as $M_{\rm WD}$ and
$M_{\rm env}$, is a crucial parameter to the nucleosynthesis
results. For smaller $X_{\rm WD}$, the explosion is less violent and
thus the cooling timescale is longer, because of the smaller nuclear
fuel. As a result, the models with low $X_{\rm WD}$ (case~A) produce
more sulfur but less oxygen than those with high $X_{\rm WD}$ (cases~A
and B). The former case is unfavorable for the production of $^7$Be,
$^{22}$Na, and $^{26}$Al.

Comparison of our nucleosynthesis results with observational abundance
estimates enables us to constrain the model parameters ($M_{\rm WD}$,
$M_{\rm env}$) for the observed ONeMg novae. We have found that the
white dwarf masses of at least four of the observed six ONeMg novae are
as low as $\simeq 1.1 M_\odot$. This is significantly smaller than the
prediction of $M_{\rm WD} \sim 1.25 - 1.35 M_\odot$ obtained by previous
hydrodynamic studies. On the other hand, our results suggest that their
envelope masses were $\gtrsim 10^{-4} M_\odot$ which are consistent with
the observational estimates of their ejected masses. In addition, the
observed fast ejection velocities for these novae ($\gtrsim
1000$~km~s$^{-1}$) are also obtained in those models. There remains a
discrepancy between these high ejected masses and those estimated by
previous hydrodynamic studies. However, a low mass white dwarf ($M_{\rm
WD} \simeq 1.1 M_\odot$) may be able to accumulate such a massive
envelope with a small mass accretion rate and a low surface temperature
(Starrfield et al. 1998).

Our results also show that the models $M_{\rm WD} \simeq 1.1 M_\odot$
with $M_{\rm env} \gtrsim 10^{-4} M_\odot$, which are the possible
explanations for most of the observed ONeMg novae,
produce significant amounts of
$^7$Be, $^{22}$Na, and $^{26}$Al:\\
1. The $\gamma$-ray line flux from the $^7$Be electron
capture is too weak to be detected with the CGRO sensitivity for the
typical distance from the sun. However, a nearby ONeMg nova could emit
the $\gamma$-rays detectable by INTEGRAL in the near future.\\
2. The mass of $^{22}$Na per event is significantly high in the models
$M_{\rm env} \gtrsim 10^{-4} M_\odot$. V1974 Cyg may have produced an
interesting amount of $^{22}$Na which is near the upper limit to the
COMPTEL sensitivity. Furthermore, we suggest that at least four ONeMg
novae in the past twenty years have produced enough $^{22}$Na to the
INTEGRAL sensitivity. The next ONeMg nova will be a promising target for
the detection of the $\gamma$-ray emitter $^{22}$Na.\\
3. The mass of $^{26}$Al per event is also significantly high in the 
models $M_{\rm env}\gtrsim 10^{-4} M_\odot$. The mass of
the Galactic $^{26}$Al which originates from ONeMg novae is estimated
to be $\lesssim 3 M_\odot$, being consistent to the COMPTEL result.
They may not be, however, major contributors according to the
$\gamma$-ray survey at 1.8~MeV by COMPTEL. The $\gamma$-ray line survey
by INTEGRAL will significantly constrain the ranges of ($M_{\rm WD}$,
$M_{\rm env}$) for ONeMg novae.

We should emphasize that hydrodynamic investigations including
multi-dimensional calculations, especially with a massive envelope, are
necessary to prove our conclusions with the one-zone approximation.
There are also other observables besides the abundances which cannot be
dealt with in this study (e.g., the surface luminosity and the ejecta
masses). Nevertheless, our results afford some new perspectives on the
future nova modelings. The future INTEGRAL survey for the $\gamma$-ray
emitters, together with abundance analyses by ultraviolet, optical, and
near infrared spectroscopies, will also impose a severe constraint on
the current nova models.

\acknowledgments

We would like to acknowledge useful discussions with T. Kajino,
S. Kubono, I. Hachisu, and J. W. Truran. We would like to express
sincere appreciation to F. -K. Thielemann for providing the data of
nuclear reaction rates. This work has been supported in part by the
grant-in-Aid for Scientific Research (05242102, 06233101) and COE
research (07CE2002) of the Ministry of Education, Science, and Culture
in Japan, and from Japan Society for Promotion of Science.

%\clearpage

\clearpage

\figcaption[f1.eps]{The ($M_{\rm WD}$, $M_{\rm env}$) sequences at which
our numerical calculations have been carried out. The dots denote this
work, and squares, triangles, and stars are, respectively, from Politano
et al. (1995), Starrfield et al. (1998), and Jos\'e and Hernanz
(1998). \label{fig1}}

\figcaption[f2.eps]{Contours of the proper values for the pressure
($P_{\rm b}/f_{\rm b}$) and the density ($\rho_{\rm b}/V_{\rm b}f_{\rm
b}$) in the logarithmic scale in the $M_{\rm WD}$--$M_{\rm env}$
space. \label{fig2}}

\figcaption[f3.eps]{The $M$--$R$ relations for various white dwarfs. The
solid line is for ${\rm O} : {\rm Ne} = 5:3$ (this work), dotted for the
completely degenerate electron gas by Chandrasekhar's method ($Y_{\rm e}
= 0.5$), broken and dot-dash for carbon and magnesium (Hamada \&
Salpeter 1961). The dots on the lines for O + Ne, carbon, and magnesium
denote at which neutronization occurs. The triangles are taken from
PSTWS95 and STWS98. \label{fig3}}

\figcaption[f4.eps]{The ratios of our nucleosynthesis results to STWS98
(sequence 6). The dots and triangles denote isotopes and elements,
respectively. \label{fig4}}

\figcaption[f5.eps]{The nucleosynthesis results for several ($M_{\rm
WD}/M_\odot$, $M_{\rm env}/M_\odot$) sequences in the $N$--$Z$
plane. The size of a circle indicates the yield at the final stage, and
the length of an arrow the net nuclear flow in the logarithmic
scale. The initial compositions are shown by dotted
circles. \label{fig5}}

\figcaption[f6.eps]{Contours of the peak temperatures at the base, the
cooling timescales, the energy generation rates per unit mass, and the
ejection velocities in the $M_{\rm WD}$--$M_{\rm env}$ space
(case~B). \label{fig6}}

\figcaption[f7.eps]{The time variations of the temperature at the base
and the nuclear energy generation rate per unit mass for ($M_{\rm
WD}/M_\odot$, $M_{\rm env}/M_\odot$) = (1.15, $10^{-4.0}$) and (1.35,
$10^{-5.5}$). \label{fig7}}

\figcaption[f8.eps]{Contours of the abundances of $^7$Be and $^{11}$B in
the logarithmic scale (case~B). \label{fig8}}

\figcaption[f9.eps]{Same as Figure~8, but for carbon, nitrogen, and
their isotope ratios. \label{fig9}}

\figcaption[f10.eps]{Same as Figure~8, but for oxygen, its isotope
ratios, and fluorine. \label{fig10}}

\figcaption[f11.eps]{Same as Figure~8, but for neon, sodium, their
isotope ratios, and $^{22}$Na. \label{fig11}}

\figcaption[f12.eps]{Same as Figure~8, but for magnesium, aluminum,
their isotope ratios, and $^{26}$Al. \label{fig12}}

\figcaption[f13.eps]{Same as Figure~8, but for silicon, its isotope
ratios, and phosphorus. \label{fig13}}

\figcaption[f14.eps]{Same as Figure~8, but for sulfur and its isotope
ratios, and the sum of chlorine, argon, potassium, and
calcium. \label{fig14}}

\figcaption[f15.eps]{Contours of the peak temperatures at the base and
the cooling timescales in the $M_{\rm WD}$--$M_{\rm env}$ space for
case~A and C. \label{fig15}}

\figcaption[f16.eps]{Same as Figure~15, but for the energy generation
rates per unit mass and the ejection velocities. \label{fig16}}

\figcaption[f17.eps]{The nucleosynthesis results for ($M_{\rm
WD}/M_\odot$, $M_{\rm env}/M_\odot$) = (1.05, $10^{-4.0}$) in the
$N$--$Z$ plane for case~A and C. \label{fig17}}

\figcaption[f18.eps]{Contours of the abundances of carbon and oxygen in
the logarithmic scale for case~A and C. \label{fig18}}

\figcaption[f19.eps]{Same as Figure~18, but for magnesium, silicon, and
sulfur. \label{fig19}}

\figcaption[f20.eps]{Same as Figure~18, but for $^7$Be, $^{22}$Na, and
$^{26}$Al. \label{fig20}}

\figcaption[f21.eps]{The ($M_{\rm WD}$, $M_{\rm env}$) sequences which
are in agreement with recent ONeMg novae, within the factor of three for
V693 CrA (triangles), V351 Pup (asterisks), and V1974 Cyg (stars), and
of five for QU Vul (circles) and V838 Her (squares). The thick signs are
the best sequences in our results. \label{fig21}}

\figcaption[f22.eps]{The ratios of our results to observational
abundance estimates. The symbols are the same as Figure~20.
\label{fig22}}

\figcaption[f23.eps]{Contours of the masses of $^7$Be, $^{22}$Na, and
$^{26}$Al per event in the logarithmic scale
(case~B). \label{fig23}}


\bigskip
\bigskip
\bigskip
\bigskip
\bigskip
\bigskip
\bigskip
\bigskip
\small

\begin{thebibliography}{}

\bibitem[Anders \& Grevesse (1989)]{Ande89}
 Anders, E., \& Grevesse, N. 1989, \gca, 53, 197
\bibitem[Andre\"a, Drechsel, \& Starrfield 1994]{Andr94}
 Andre\"a, J., Drechsel, H., \& Starrfield, S. 1994, \aap, 291, 869
\bibitem[Austin et al. 1996]{Aust96}
 Austin, S. J., Wagner, R. M., Starrfield, S., Shore, S. N., Sonneborn, G., 
 \& Bertram, R. 1996, \aj, 111, 869
\bibitem[Boffin et al. 1993]{Boff93}
 Boffin, H. M. J., Paulus, G., Arnould, M., \& Mowlavi, N. 1993, \aap, 279, 173
\bibitem[Caughlam \& Fowler 1988]{Caug88}
 Caughlaln, G. R., \& Fowler, W. A. 1988, Atomic Data Nucl. Data Tables, 40
\bibitem[Gehrz et al. 1998]{Gehr98} Gehrz, R. G., Truran, J. W.,
 Williams, R. E., \& Starrfield, S. 1998, \pasp, 110, 3
\bibitem[Champagne et al. 1988]{Cham88}
 Champagne, A. E., Cella, C. H., Kouzes, R. T., Lowry, M. M., Magnus, P.V., 
 Smith, M. S., \& Mao, Z. Q. 1988, \nphysa, 487, 433
\bibitem[Champagne, Brown, \& Sherr 1993]{Cham93}
 Champagne, A. E., Brown, B. A., \& Sherr, R. 1993, \nphysa, 556, 123
\bibitem[Chochol et al. 1997]{Choc97}
 Chochol, D., Grygar, J., Pribulla, T., Kom\v z\'ik, R., Hric, L., 
 \& Elkin, V. 1997, \aap, 318, 908
\bibitem[Coc et al. 1995]{Coc95}
 Coc, A., Mochkovitch, R., Oberto, Y., Thibaud, J. P., \& Vangioni-Flam, E. 
 1995, \aap, 299, 479
\bibitem[D'Antona \& Matteucci 1991]{Dant91}
 D'Antona, F. \& Matteucci, F. 1991, \aap, 248, 62
\bibitem[Diehl et al. 1994]{Dieh94}
 Diehl, R., Dupraz, C., Bennet, K., Bloemen, H., deDoer, H., 
 Hermsen, W., Lichti, G. G., McConnell, M., Morris, D., Ryan, J., 
 Sch\"onfelder, V., Steinle, H., Strong, A. W., Swanenburg, B. N., 
 Varendorff, M., \& Winkler, C. 1994, \apjs, 92, 429
\bibitem[Diehl et al. 1995]{Dieh95}
 Diehl, R., Dupraz, C., Bennet, K., Bloemen, H., Hermsen, W., 
 Kn\"odlseder, J., Lichti, G., Morris, D., Ryan, J., Sch\"onfelder, V., 
 Steinle, H., Strong, A. W., Swanenburg, B. N., Varendorff, M., 
 \& Winkler, C. 1995, \aap, 298, 445
\bibitem[Forestini, Paulus, \& Arnould 1991]{Fore91}
 Forestini, M., Paulus, G., \& Arnould, M. 1991, \aap, 252, 597
\bibitem[Fujimoto 1982a]{Fuji82a}
 Fujimoto, M. Y. 1982a, \apj, 257, 752
\bibitem[Fujimoto 1982b]{Fuji82b}
 ----------. 1982b, \apj, 257, 767
\bibitem[Gehrz et al. 1994]{Gehr94}
 Gehrz, R. D., Woodward, C. E., Greenhouse, M. A., Starrfield, S., 
 Wooden, D. H., Witteborn, F. C., Sandford, S. A., Allamandola, L. J., 
 \& Bregman, J. D. 1994, \apj, 421, 762
\bibitem[Glasner, Livne, \& Truran 1997]{Glas97}
 Glasner, S. A., Livne, E., \& Truran, J. W. 1997, \apj, 475, 754
\bibitem[G\'omez-Gomar et al. 1998]{Gome98}
 G\'omez-Gomar, J., Hernanz, M., Jos\'e, J., \& Isern, J 1998, \mnras,
 296, 913
\bibitem[Greenhouse et al. 1988]{Gree88}
 Greenhouse, M. A., Grasdalen, G. L., Hayward, T. L., Gehrz, R., D., 
 \& Jones, T. J. 1988, \aj, 95, 172
\bibitem[Hamada \& Salpeter 1961]{Hama61}
 Hamada, T. \& Salpeter, E. E. 1961, \apj, 134, 683
\bibitem[Hashimoto, Iwamoto, \& Nomoto (1993)]{Hash93}
 Hashimoto, M., Iwamoto, K., \& Nomoto, K. 1993, \apjl, 414, 105
\bibitem[Hatano et al. 1997]{Hata97}
 Hatano, K., Branch, D., Fisher, A. \& Starrfield, S. 1997, \mnras, 290, 113
\bibitem[Hayward et al. 1996]{Hayw96}
 Hayward, T. L., Saizar, P., Gehrz, R. D., Benjamin, R. A., Mason, C. G., 
 Houck, J. R., Miles, J. W., Gull, G. E., \& Schoenwald, J. 1996, \apj, 
 469, 854
\bibitem[Hernanz et al. 1996]{Hern96}
 Hernanz, M., Jos\'e, J., Coc, A., \& Isern, J. 1996, \apjl, 465, 27
\bibitem[Herndl et al. 1995]{Hern95}
 Herndl, H., G\"orres, J., Wiescher, M., Brown, B. A., \& Van Wormer 1995, 
 \prc, 52, 1078
\bibitem[Hillebrandt \& Thielemann 1982]{Hill82}
 Hillebrandt, W. \& Thielemann, F. -K. 1982, \apj, 255, 617
\bibitem[Iben \& Tutukov 1985]{Iben85}
 Iben, I. Jr., \& Tutukov, A. V. 1985, \apjs, 58, 661
\bibitem[Iben, Fujimoto, \& MacDonald 1991]{Iben91}
 Iben, I. Jr., Fujimoto, M. Y., \& MacDonald, J. 1991, \apjl, 375, 27
\bibitem[Iben \& Tutukov 1996]{Iben96}
 Iben, I. Jr., \& Tutukov, A. V. 1996, \apjs, 105, 145
\bibitem[Ichimaru \& Kitamura 1994]{Ichi94}
 Ichimaru, S. \& Kitamura, H. 1994, in Proc. Int. Semi. 
 on Elementary Processes in Dense Plasmas, ed. S. Ichimaru \& S. Ogata 
 (Addison-Wesley), 113
\bibitem[Iglesias \& Rogers 1993]{Igle93}
 Iglesias, C. A., \& Rogers, F. J. 1993, \apj, 412, 752
\bibitem[Iliadis et al. 1996]{Ilia96}
 Iliadis, C., Buchmann, L., Endt, P. M., Herndl, H., \& Wiescher, M. 
 1996, \prc, 53, 475
\bibitem[Iyudin et al. 1995]{Iyud95}
 Iyudin, A. F., Bennett, K., Bloemen, H., Diehl, R., Hermsen, W., 
 Lichti, G., Morris, D, Ryan, J., Sch\"ondelder, V., Steinle, H., 
 Strong, A., Varendorff, M., \& Winkler, C. 1995, \aap, 300, 422
\bibitem[Jos\'e, Hernanz, \& Coc 1997]{Jose97}
 Jos\'e, J., Hernanz, M., \& Coc, A. 1997, \apjl, 479, 55
\bibitem[Jos\'e \& Hernanz 1998]{Jose98}
 Jos\'e, J., \& Hernanz, M. 1998, \apj, 494, 680
\bibitem[Kercek, Hillebrandt, \& Truran 1998a]{Kerc98}
 Kercek, A., Hillebrandt, W., \& Truran, J. W. 1998a, \aap, 337, 379
\bibitem[Kercek, Hillebrandt, \& Truran 1998b]{Kerc98b}
 ----------. 1998b, \aap, submitted
\bibitem[Kolb \& Politano 1997]{Kolb97}
 Kolb, U. \& Politano, M. 1997, \aap, 319, 909
\bibitem[Kovetz \& Prialnik 1997]{Kove97}
 Kovetz, A. \& Prialnik, D. 1997, \apj, 477, 356
\bibitem[Krautter et al. 1996]{Krau96}
 Krautter, J., \"Ogelman, H., Starrfield, S., Wichmann, R., 
 \& Pfeffermann, E. 1996, \apj, 456, 788
\bibitem[Kubono et al. 1994]{Kubo94}
 Kubono, S., Yun, C. C., Boyd, R. N., Buchmann, L. R., Fuchi,Y., 
 Hosaka, M., Ikeda, N., Jiang, C. L., Katayama, I., Kawashima, H.,
 Miyatake, H., Niizeki, T., Nomura, T., Odahara, A., Ohura, M.,
 H. Orihara, H., Rolfs, C., Shimoda, T., Tajima, Y., Tanaka, M. H.,
 \& Toyokawa, H. 1994, Z. Phys. A, 348, 59
\bibitem[Kubono et al. 1996]{Kubo96}
 Kubono, S., Hosaka, M., Strasser, P., Guimaraes, V., Jeong,
 S. C., Katayama, I., Miyachi, T., Nomura, T., Tanaka, M. H., Fuchi,
 Y., Kawashima, H., Kato, S., Yun, C. C., Orihara, H., Niizeki, T.,
 Miyatake, H., Shimoda, T., Kajino, T., and Wanajo, S. 1997, 
 \nphysa, 621, 195
\bibitem[Kutter \& Sparks 1987]{Kutt87}
 Kutter, G. S., \& Sparks, W. M. 1987, \apj, 321, 386
\bibitem[Livio \& Truran 1994]{Livi94}
 Livio, M., \& Truran, J. W. 1994, \apj, 425, 797
\bibitem[MacDonald 1983]{MacD83}
 MacDonald, J. 1983, \apj, 267, 732
\bibitem[Mahoney et al. 1984]{Maho84}
 Mahoney, W. A., Ling, J. C.,  Wheaton, W. A., \& Jacobson, A. S. 1984, \apj, 
 286, 578
\bibitem[Matteucci, D'Antona, \& Timmes 1995]{Matt95}
 Matteucci, F., D'Antona, F., \& Timmes, F. X. 1995, \aap, 303, 460
\bibitem[Meynet et al. 1997]{Meyn97}
 Meynet, G., Arnould, M., Prantzos, N., \& Paulus, G. 1997, \aap, 320, 460
\bibitem[Nofar, Shaviv, \& Starrfield 1991]{Nofa91}
 Nofar, I., Shaviv, G., \& Starrfield, S. 1991, \apj, 369, 440
\bibitem[Nomoto 1984]{Nomo84}
 Nomoto, K. 1984, \apj, 277, 291
\bibitem[Nomoto 1987]{Nomo87}
 ----------. 1987, \apj, 322, 206
\bibitem[Nomoto \& Hashimoto 1988]{Nomo88}
 Nomoto, K., \& Hashimoto, M.  1988, \physrep, 163, 13
\bibitem[Politano et al. 1995]{Poli95}
 Politano, M., Starrfield, S., Truran, J. W., Weiss, A., \& Sparks, W. M. 
 1995, \apj, 448, 807
\bibitem[Paresce et al. 1995]{Pare95}
 Paresce, F., Livio, M., Hack, W., \& Korista, K. 1995, \aap, 299, 823
\bibitem[Pavelin et al. 1993]{Pave93}
 Pavelin, P. E., Davis, R. J., Morrison, L. V., Bode, M. F., \& Ivison, R. J. 
 1993, \nat, 363, 424
\bibitem[Prantzos \& Cass\'e 1986]{Pran86}
 Prantzos, N., \& Cass\'e, M. 1986, \apj, 307, 324
\bibitem[Prantzos 1993]{Pran93}
 Prantzos, N. 1993, \apj, 405, L55
\bibitem[Prantzos \& Diehl 1996]{Pran96a}
 Prantzos, N., \& Diehl, R.  1996, \physrep, 267, 1
\bibitem[Prantzos 1996]{Pran96b}
 Prantzos, N. 1996, \aaps, 120, 303
\bibitem[Prialnik \& Kovetz 1984]{Pria84}
 Prialnik, D., \& Kovetz, A. 1984, \apj, 281, 367
\bibitem[Prialnik \& Kovetz 1995]{Pria95}
 Prialnik, D., \& Kovetz, A. 1995, \apj, 445, 789
\bibitem[Retter, Leibowitz, \& Ofek 1997]{Rett97}
 Retter, A., Leibowitz, E. M., \& Ofek, E. O. 1997, \mnras, 286, 745
\bibitem[Ritossa, Garc\'{\i}a, \& Iben 1996]{Rito96}
 Ritossa, C., Garc\'{\i}a, E., -B., \& Iben, I. 1996, \apj, 460, 489
\bibitem[Ritter et al. 1991]{Ritt91}
 Ritter, H., Politano, M., Livio, M., \& Webbink, R. F. 1991, \apj, 376, 177
\bibitem[Saizar et al. 1992]{Saiz92}
 Saizar, P., Starrfield, S., Ferland, G. J., Wagner, R. M., Truran, J. W., 
 Kenyon, S. J., Sparks, W. M., Williams, R. E., \& Stryker, L. L. 1992, \apj, 
 398, 651
\bibitem[Saizar et al. 1996]{Saiz96}
 Saizar, P., Pachoulakis, I., Shore, S. N., Starrfield, S., 
 Williams, R. E., Rothschild, E., \& Sonneborn, G. 1996, \mnras, 
 279, 280
\bibitem[Schmidt et al. 1995]{Schm95}
 Schmidt, S., Rolfs, C., Schulte, W. H., Trautvetter, H. P., Kavanagh, R. W., 
 Hategan, C., Faber, S., Valnion, B. D., \& Graw, G. 1995, \nphysa, 591, 227
\bibitem[Shafter 1997]{Shaf97}
 Shafter, A. W. 1997, \apj, 487, 226
\bibitem[Shields 1996]{Shie96}
 Shields, G. A. 1996, \apjl, 461, 9
\bibitem[Shore et al. 1993]{Shor93}
 Shore, S. N., Sonneborn, G., Starrfield, S., Gonzalez-Riestra, R., 
 \& Ake, T. B. 1993, \aj, 106, 2408
\bibitem[Sion et al. 1997]{Sion97}
 Sion, E. M., Cheng, F. H., Sparks, W. M., Szkody, P., Huang, M., 
 \& Hubeny, I. 1997, \apjl, 480, 17
\bibitem[Snijders et al. 1987]{Snij87}
 Snijders, M. A. J., Batt, T. J., Roche, P. F., Seaton, M. J., Morton, 
 D. C., Spoelstra, T. A. T., \& Blades, J. C. 1987, \mnras, 228, 329
\bibitem[Starrfield, Truran, \& Sparks 1978]{Star78}
 Starrfield, S., Truran, J. W., \& Sparks, W. M. 1978, \apj, 226, 186
\bibitem[Starrfield, Sparks, \& Shaviv 1988]{Star88}
 Starrfield, S., Sparks, W. M., \& Shaviv, G. 1988, \apjl, 325, 35
\bibitem[Starrfield et al. 1993]{Star93}
 Starrfield, S., Truran, J. W., Politano, M., Sparks, W. M., Nofar, I., 
 \& Shaviv, G. 1993, \physrep, 227, 223
\bibitem[Starrfield et al. 1998]{Star98}
 Starrfield, S., Truran, J. W., Wiescher, M. C., Sparks, W. M. 1998, 
 \mnras, 296, 502
\bibitem[Sugimoto \& Fujimoto 1978]{Sugi78}
 Sugimoto, D., \& Fujimoto, M. Y. 1978, \pasj, 30, 467
\bibitem[Taylor et al. 1987]{Tayl87}
 Taylor, A. R., Seaquist, E. R., Hollis, J. M., \& Pottasch, S. R. 1987, 
 \aap, 183, 38
\bibitem[Timmes et al. 1995]{Timm95}
 Timmes, F. X., Woosley, S. E., Hartmann, D. H., \& Hoffman, R. D. 1995, 
 \apj, 449, 204
\bibitem[Timmermann et al. 1988]{Timm88}
 Timmernamm, R., Becker, H. W., Rolfs, C., Schr\"oder, U., 
 \& Trautvetter, H. P. 1988, \nphysa, 477, 105
\bibitem[Thielemann et al. 1994]{Thie94}
 Thielemann, F.-K., Kratz, K.-L., Pfeiffer, B., Rauscher, T., 
 van Wormer, L., \& Wiescher, M. C. 1994, \nphysa, 570, 329
\bibitem[Thielemann 1995]{Thie95}
 Thielemann, F.-K. 1995, private communication
\bibitem[Truran et al. 1977]{Trur77}
 Truran, J. W., Starrfield, S., Strittmatter, P. A., Wyatt, S. P., 
 \& Sparks, W. M. 1977, \apj, 211, 539
\bibitem[Truran 1982]{Trur82}
 Truran, J. W. 1982, Essays in Nuclear Astrophysics, 
 ed. C. A. Barnes, D. D. Clayton, \& D. N. Schramm 
 (Cambridge: Cambridge University Press), p. 467
\bibitem[Truran et al. 1987]{Trur87}
 Truran, J. W., Thielemann, F.-K., \& Arnould, M. 1987, 
 Nuclear Astrophysics, ed. Hillebrandt (Springer), p. 91
\bibitem[Vanlandingham et al. 1996]{Vanl96}
 Vanlandingham, K. M., Starrfield, S., Wagner, R. M., Shore, S. N., 
 \& Sonneborn, G. 1996, \mnras, 282, 563
\bibitem[Vanlandingham, Starrfield, \& Shore 1997]{Vanl97}
 Vanlandingham, K. M., Starrfield, S., \& Shore, S. N. 1997, \mnras, 290, 87
\bibitem[Van Wormer et al. 1994]{VanW94}
 Van Wormer, L., G\"orres, J., Iliadis, C., Wiescher, M., 
 \& Thielemann, F.-K. 1994, \apj, 432, 326
\bibitem[Wagner \& DePoy 1996]{Wagn96}
 Wagner, R. M., \& DePoy, D. L. 1996, \aap, 218, 123
\bibitem[Walter \& Maeder 1989]{Walt89}
 Walter, R., \& Maeder, A. 1989, \aap, 218, 123
\bibitem[Wanajo et al. 1997a]{Wana97a}
 Wanajo, S., Nomoto, K., Hashimoto, M., Kajino, T., \& Kubono, S. 1997, 
 \nphysa, 616, 91
\bibitem[Wanajo et al. 1997b]{Wana97b}
 Wanajo, S., Nomoto, K., Truran, J. W., \& Hashimoto, M. 1997, 
 \nphysa, 621, 499
\bibitem[Wanajo, Ishimaru, \& Kajino 1998]{Wana98}
 Wanajo, S., Ishimaru, Y., \& Kajino, T. 1998, in
 Origin of Matter And Evolution of Galaxies 97, ed. S. Kubono,
 T. Kajino, K. Nomoto, \& I. Tanihata (World Scientific), 262
\bibitem[Ward \& Fowler 1980]{Ward80}
 Ward, R. A., \& Fowler, W. A. 1980, \apj, 238, 266
\bibitem[Weiss \& Truran 1990]{Weis90}
 Weiss, A., \& Truran, J. W. 1990, \aap, 238, 178
\bibitem[Wiescher et al. 1986]{Wies86}
 Wiescher, M., G\"orres, J., Thielemann, F.-K., \& Ritter, H. 1986, 
 \aap, 160, 56
\bibitem[Williams et al. 1985]{Will85}
 Williams, R. E., Ney, E. P., Sparks, W. M., Starrfield, S. G., Wyckoff, 
 S., \& Truran, J. W. 1985, \mnras, 212, 753
\bibitem[Woodward et al. 1992]{Wood92}
 Woodward, C. E., Gehrz, R. D., Jones, T. J., \& Lawrence, G. F. 1992, 
 \apjl, 384, 41
\bibitem[Woodward et al. 1995]{Wood95}
 Woodward, C. E., Greenhouse, M. A., Gehrz, R. D., Pendleton, Y. J., 
 Joyce, R. R., Van Buren, D., Fischer, J., Jennerjohn, N. J., 
 \& Kaminski, C. D. 1995, \apj, 438, 921
\bibitem[Woodward et al. 1997]{Wood97}
 Woodward, C. E., Gehrz, R. D., Jones, T. J., Lawrence, G. F., 
 \& Skrutskie, M. F. 1997, \apj, 477, 817
\bibitem[Yungelson, Livio, \& Tutukov 1997]{Yung97}
 Yungelson, L., Livio, M., \& Tutukov, A. 1997, \apj, 481, 127

\end{thebibliography}
\end{document}